\documentclass[a4paper,11pt, twocoloumn]{article}
\usepackage{amsmath, amssymb}
\usepackage{array}
\usepackage[english]{babel}
\usepackage{fancyhdr}
\usepackage[T1]{fontenc}
\usepackage[margin=1in]{geometry}
\usepackage{graphicx}
\usepackage{siunitx}
\usepackage[utf8]{inputenc}
\usepackage{lastpage}
\usepackage{longtable}
\usepackage{mathtools, slashed}
\usepackage{multicol}
\usepackage{multirow}
\usepackage{subfigure}
\usepackage{tabulary}
\usepackage{tabularx}
\usepackage{verbatim}
\usepackage{enumitem}
\usepackage{capt-of}
\usepackage{xcolor}
\setlist[enumerate]{noitemsep}
\setlist[itemize]{noitemsep}
\setlist[description]{noitemsep}
\usepackage[margin=10pt, font=small, labelfont=bf]{caption}
\usepackage{longtable}
\usepackage{nameref}
\usepackage{hyperref}

\begin{document}
	
\renewcommand{\subsectionmark}[1]{\markboth{#1}{}}

\begin{center}
	{\Large 
		\textbf{
		Giant light emission enhancement in strain-engineered InSe/MS$_2$ (M=Mo,W) van der Waals heterostructures
		}}
\end{center}

\vspace{1 mm}

\begin{center}
	{\large Elena Blundo,$^{1,*}$ Marzia Cuccu,$^{1}$ Federico Tuzi,$^{1}$ Michele Re Fiorentin,$^{2}$ Giorgio Pettinari,$^{3}$ Atanu Patra,$^{1}$ Salvatore Cianci,$^{1}$ Zakhar Kudrynskyi,$^{4}$ Marco Felici,$^{1}$ Takashi Taniguchi,$^{5}$ Kenji Watanabe,$^{6}$ Amalia Patan\`{e},$^{8}$ Maurizia Palummo,$^{6}$ and Antonio Polimeni$^{1,*}$}
\end{center}

\begin{center}
	\textit{\mbox{}$^1$ Physics Department, Sapienza University of Rome, Piazzale Aldo Moro 5, 00185 Rome, Italy.\\
	\mbox{}$^2$ Department of Applied Science and Technology, Politecnico di Torino, corso Duca degli Abruzzi 24, 10129 Torino, Italy\\
	\mbox{}$^3$ Institute for Photonics and Nanotechnologies, National Research Council, 00133 Rome, Italy.\\
	\mbox{}$^4$ Faculty of Engineering, University of Nottingham, Nottingham, NG7 2RD, UK.\\
	\mbox{}$^5$ Research Center for Materials Nanoarchitectonics, National Institute for Materials Science,  1-1 Namiki, Tsukuba 305-0044, Japan.\\
        \mbox{}$^6$ Research Center for Electronic and Optical Materials, National Institute for Materials Science, 1-1 Namiki, Tsukuba 305-0044, Japan.\\
        \mbox{}$^7$ School of Physics and Astronomy, University of Nottingham, Nottingham, NG7 2RD, UK.\\
	\mbox{}$^8$ INFN, Dipartimento di Fisica, Universitá di Roma Tor Vergata, Via della Ricerca Scientifica 1, 00133 Rome, Italy.\\
	}
\end{center}

\mbox{}$^*$ Corresponding authors: antonio.polimeni@uniroma1.it, elena.blundo@uniroma1.it

\begin{center}
    {\small 30 August 2024}
\end{center}

\date{}

\vspace{10 mm}

\noindent
\textbf{Abstract}\\
\mbox{}\\
Two-dimensional crystals stack together through weak van der Waals (vdW) forces, offering unlimited possibilities to play with layer number, order and twist angle in vdW heterostructures (HSs). The realisation of high-performance optoelectronic devices, however, requires the achievement of specific band alignments, $k$-space matching between conduction band minima and valence band maxima, as well as efficient charge transfer between the constituent layers. Fine tuning mechanisms to design ideal HSs are lacking. Here, we show that layer-selective strain engineering can be exploited as an extra degree of freedom in vdW HSs to tailor their band alignment and optical properties. To that end, strain is selectively applied to MS$_2$ (M=Mo,W) monolayers in InSe/MS$_2$ HSs. 
This triggers 
a giant PL enhancement of the highly tuneable but weakly emitting InSe by one to three orders of magnitude. 
Resonant PL excitation measurements, supported by first-principle calculations, provide evidence of a strain-activated direct charge transfer from the MS$_2$ MLs toward InSe.
This significant emission enhancement achieved for InSe widens its range of applications for optoelectronics.



\twocolumn

\section*{Introduction}
van der Waals (vdW) heterostructures (HSs) offer a vast playground for the realisation of novel electronic and optoelectronic devices, due to the multitude of degrees of freedom they display, such as layer number, order and twist angle. The weak vdW adhesion \cite{blundo_adhesion_prl} that keeps together different two-dimensional (2D) crystals is responsible for this unprecedented tunability, overcoming lattice mismatch issues and rotational constraints \cite{Heterostructures_GeimGriogorieva}. 
This high tunability has been exploited to engender novel phenomena ---such as superconductivity in twisted multilayer graphene \cite{HeterostructuresMagicAngle_Cao}--- and to realise efficient electronic devices \cite{HeterostructuresFET_Britnell,HeterostructuresFET_Georgiou,HeterostructuresTunnelFET_Britnell,HeterostructuresLED_Pospischil,HeterostructuresLED_Baugher,HeterostructuresLED_Ross}. However, the achievement of high-performance optoelectronic devices remains a challenge due to the necessity to find materials with specific band alignments, $k$-space matching conduction band minima (CBM) and valence band maxima (VBM), and efficient charge transfer between different layers \cite{moire_potential_HS_brem,Blundo_planar_HS_nc}. Fine tuning mechanisms to design ideal HSs are still missing. Strain has been used to shift the photoluminescence (PL) signal of HSs made of transition metal dichalcogenides (TMDs) \cite{Heterostructures_strain_IX_cho} and to modify the geometry of the moir\'{e} potential in twisted bilayers \cite{Heterostructures_strain_moire_bai}. In all cases, the entire HS was stretched. 
Here, we propose a novel paradigm, by selectively straining only one of the constituent materials of vdW HSs formed by MS$_2$ (M=Mo,W) TMD monolayers (MLs) and InSe thin flakes. Specifically, strain is applied to MS$_2$ MLs: first, to achieve hybridised K/$\Gamma$ VB states being in quasi-resonance with the VBM of the unstrained InSe; and second, to favour a mixing between defect states in MS$_2$ and its CB states, allowing for electron tunnelling to InSe without momentum transfer. These conditions enable an efficient electronic coupling between the TMD and InSe, without any requirements on the twist angle.

The choice of InSe is grounded on the excellent properties it exhibits. Indeed, while MoS$_2$ and WS$_2$ MLs show intense light emission \cite{blundo_review}, their mobility $\mu$ (up to $\sim 200$ cm$^2$V$^{-1}$s$^{-1}$) \cite{mobility_MoS2_Yoon,mobility_MoS2_Radisavljevic} is far from being as exceptional as that of graphene, where $\mu$ exceeds $10^{4}$ cm$^2$V$^{-1}$s$^{-1}$ \cite{Review_RiseOfGr_geim}. InSe features instead excellent transport properties, with $\mu$ values exceeding $10^{3}$ cm$^2$V$^{-1}$s$^{-1}$ \cite{InSe_high_mu_adv_mat,InSe_high_mu_nat_nan}, and a quasi-direct and tunable optical bandgap $E_\mathrm{gap}$, which makes it particularly appealing $e.g.$ for fast photodetectors \cite{diodes_InSe_frisenda} operating from the UV to the near-IR range. Indeed, $E_\mathrm{gap}$ varies from about 1.2 eV to 2.0 eV, going from bulk to 2-layer (L) crystals (for MLs, the lowest-energy transition has an increased indirect character, and it is optically inactive for in-plane polarised light and only weakly coupled to z-polarised light) \cite{InSe_dir_indir,InSe_high_mu_nat_nan}.

\begin{figure}[htpb]
\includegraphics[width=7.5cm]{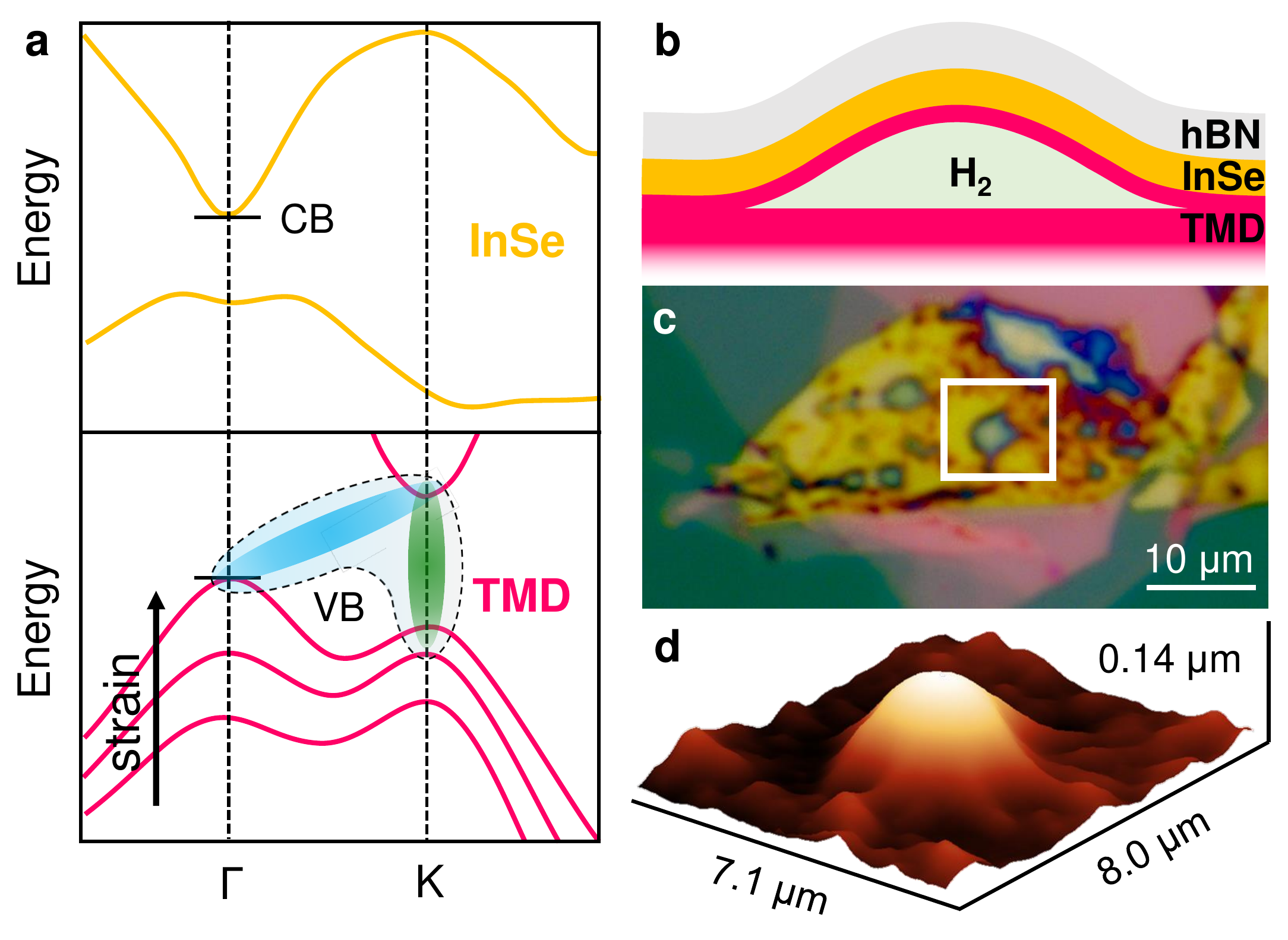}
\caption{\textbf{Heterostructured InSe/TMD bubbles.} \textbf{a} Sketch of the band structure of few-layer-thick InSe and of 1-layer-thick TMDs. The effect of strain on the VB of TMDs is highlighted: For high strains, the valley at $\Gamma$ goes above that at K and direct (green) and indirect (cyan) excitons hybridise. \textbf{b} Sketch of the system studied in this work, consisting in an heterostructured bubble: a few-layer-thick InSe flake is deposited atop of a strained TMD-ML in the shape of bubble; hBN is used to cap the system. \textbf{c} Optical image of a flake with heterostructured bubbles. \textbf{d} AFM image of the bubble within the white rectangle in panel \textbf{c}.}
\label{fig:system}
\end{figure}

Research on InSe has rapidly developed, with the fabrication of HSs, such as graphene/InSe \cite{InSe_graphene}, p-InSe/n-In$_2$O$_3$ \cite{InSe_InSe_pn}
and n-InSe/p-GaSe \cite{InSe_GaSe_pn_yan}, resulting in junctions with excellent transport characteristics. The use of InSe for optoelectronic devices, though, is hampered by its relatively low radiative efficiency. As a matter of fact, while the CBM of InSe is located at $\Gamma$, the VB has a camel's back shape, 
with the VBM slightly off the $\Gamma$ point (see Fig.\ \ref{fig:system}\textbf{a}). For a number of layers $N \gtrsim 6$,
the VBM approaches $\Gamma$ and InSe virtually becomes a direct gap semiconductor \cite{InSe_theory_band_structures}. However, the electric dipole orientation of InSe is perpendicular to the exfoliation planes that leads to a poor coupling to light directed perpendicular to the InSe plane \cite{InSe_theory_band_structures,InSe_pillars_mazumder,InSe_dipole_brotons-gisbert,InSe_lasing}, namely, the geometry mainly employed in optical devices.
Several works have reported on strategies to increase the optical efficiency of InSe ---including bending of the flakes through pillars \cite{InSe_pillars_mazumder}, ridges \cite{Shi_InSe_ridge}, and nanotexturing \cite{Brotons_nanotexturing_InSe}, or the coupling of InSe with inorganic perovskites \cite{InSe_perovskites}--- but increases of only a factor of 2-3 have been typically achieved.

In a recent work \cite{Heterostructures_InSe_IX_ubrig}, InSe was coupled to multilayer ($N \geq 2$) TMDs, whose VBM lies at the $\Gamma$ point. A type-II alignment was achieved in such HSs, with the observation of momentum-space direct (at $\Gamma$) interlayer excitons, formed by electrons at the InSe CBM and holes at the TMD VBM.  Although these HSs present the advantage of avoiding rotational constraints, their light emission is hindered by a poor optical efficiency and significant sample-to-sample fluctuations. These issues are likely ascribable to the intrinsically low radiative efficiency of InSe (as discussed above) and the \emph{k}-space indirect nature of the bandgap in TMD multilayers. Additionally, the spatially indirect nature of these interlayer excitons makes their emission detectable mainly at cryogenic temperatures.

In this work, instead, InSe is coupled to a TMD (MS$_2$) ML, whose optical bandgap is made indirect ($\Gamma_\mathrm{VB}-\mathrm{K}_\mathrm{CB}$) by strain, yet ---at variance with TMD multilayers--- it is characterised by a remarkable oscillator strength thanks to the hybridisation of direct and indirect exciton states \cite{Blundo_prr,Blundo_magneto_prl}, see Fig.\ \ref{fig:system}\textbf{a}. In fact, a fine tuning of strain leads to a unique electronic configuration for the TMD ML in which excitons with an admixed direct-indirect character are observed \cite{Blundo_magneto_prl}. In turn, the strained MS$_2$ ML retains a large light-to-charge conversion.
Here, we show that the unique band structure configuration of the HS obtained by this approach, as evaluated by Density Functional Theory (DFT) calculations, enables an efficient (strain+defect)-assisted tunneling of photo-generated electrons and holes from the strained ML toward InSe, giving rise to a giant light emission enhancement of InSe.

\section*{Results and discussion}

To create HSs in which only the MS$_2$ crystal is subject to high strain, while the InSe layer is not, we exploited strained WS$_2$ and MoS$_2$ MLs in the shape of micro-bubbles. The bubbles were created as in Ref.\ \cite{Tedeschi_AdvMater}: Bulk flakes are exposed to a low-energy ionised hydrogen beam; protons penetrate through the topmost layer and molecular hydrogen forms and accumulates leading to the formation of micro-bubbles on the flake surface. Such bubbles mainly have thickness of just 1 layer \cite{Tedeschi_AdvMater,Liu_plateau_NatCommun}, and host sizeable strains (up to more than 4\%), whose extent increases from the edge toward the centre \cite{blundo_adhesion_prl,review_bubbles_cui,blundo_hBN-bubbles_nanolett}. Optical spectroscopy and first-principle calculations revealed that the strain in the bubbles induces a direct-to-indirect transition \cite{Blundo_prr} (with the VBM shifting from K to $\Gamma$, see Fig.\ \ref{fig:system}\textbf{a}), and that the nearly-resonant direct and indirect excitons hybridise \cite{Blundo_magneto_prl}, leading to an efficient PL emission even when the indirect exciton is the lowest energy state. These properties make strained TMDs a promising system to be coupled with InSe.

\begin{figure*}[htpb]
\centering
\includegraphics[width=0.75\textwidth]{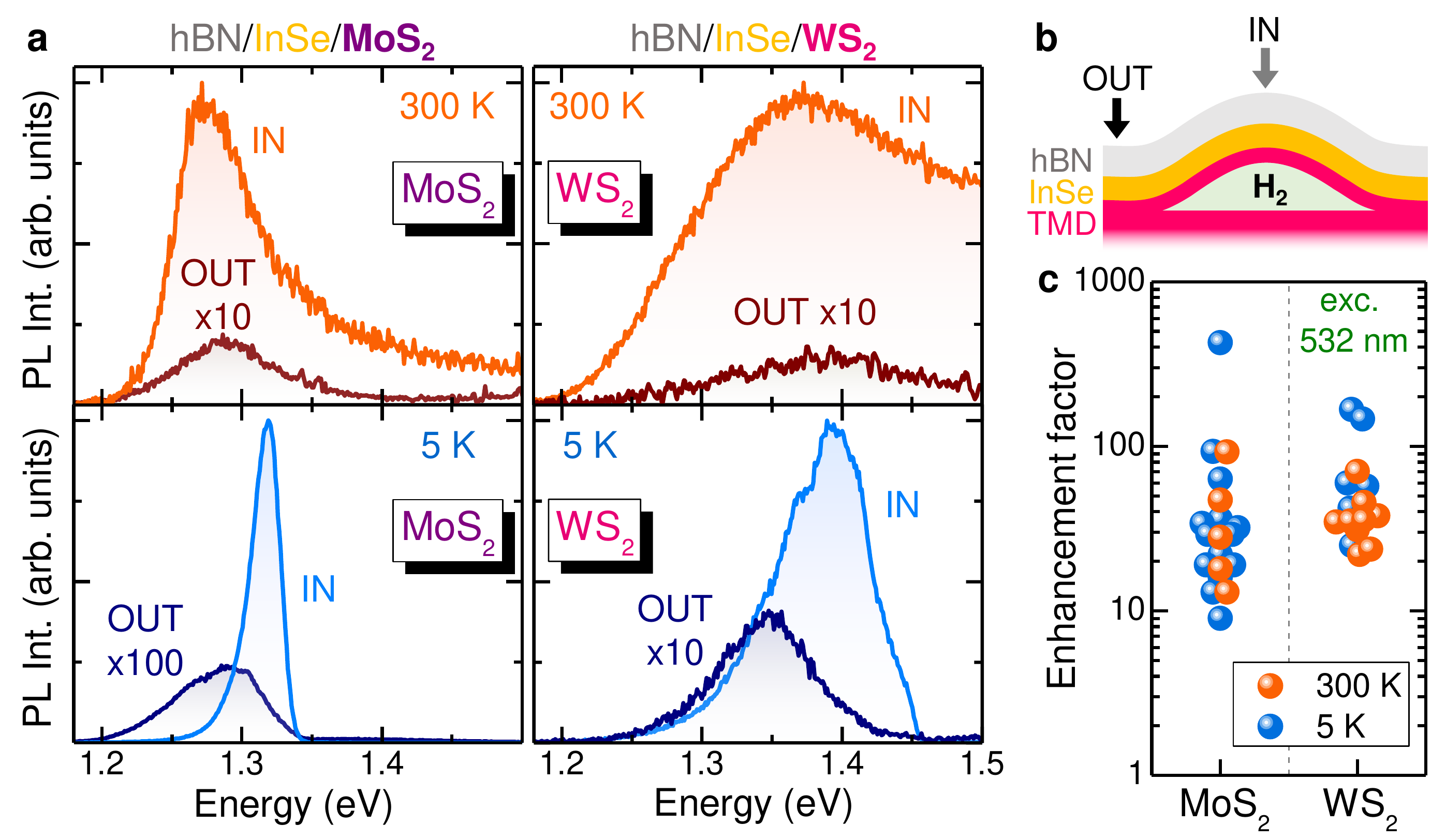}
\caption{\textbf{Giant InSe emission enhancement in selectively strained InSe/MS$_2$ heterostructures.} $\mu$-PL spectra at 300 K and at 5 K of two HS-bubbles (one with MoS$_2$ as TMD, left, and one with WS$_2$, right) and of the region right outside the bubble, as indicated in the sketch in panel (b). (c) Summary of the ratios between the PL intensity in the HS-bubbles and outside, measured in several MoS$_2$ and WS$_2$ based structures at 5 K or at 300 K.}
\label{fig:PL}
\end{figure*}
InSe and hBN flakes were mechanically exfoliated onto PDMS, and flakes with the desired thickness ($\sim$ 6-10 L for InSe, $< 20$ nm for hBN) were identified based on their optical contrast. The InSe flakes were then deposited on some selected TMD bubbles; right after, hBN was deposited atop of the HS to prevent its oxidation \cite{InSe_InSe_pn}. See \textcolor{purple}{Supporting Information, Methods} for details. A sketch of the final heterostructured bubble (HS-bubble) is shown in Fig.\ \ref{fig:system}\textbf{b}, while the optical  and 3D atomic force microscope (AFM) images of a real sample are shown in panels \textbf{c} and \textbf{d}, respectively.

The optical properties of the HS-bubbles were investigated by excitation with a 532 nm (2.33 eV) laser and by keeping the samples in vacuum to minimise sample oxidation. Micro-PL ($\mu$-PL) measurements were performed from 5 K up to room-temperature (RT). 
Interestingly, the heterostructuring process has profound implications on the mechanics of the system in the low temperature regime. In fact, hydrogen-filled TMD bubbles suddenly deflate at about 30 K due to the gas-to-liquid phase transition of H$_2$ \cite{Tedeschi_AdvMater,Cianci_SPEs_WS2_domes}, while we do not observe any major change in the morphology of our HS-bubbles even at 5 K, see \textcolor{purple}{Supporting Note 1}. As discussed therein, we attribute this to a tie-beam-like effect played by the InSe/hBN flakes, similar to that characteristic of tied-arch bridges or of the Brunelleschi's Dome.
Raman studies clearly demonstrate that the MS$_2$ MLs are characterised by biaxial strains around 2 $\%$ and that only a moderate strain reduction is observed when decreasing $T$ from RT to 5 K, see \textcolor{purple}{Supporting Note 2}.

$\mu$-PL measurements on single HS-bubbles, both at RT and low \emph{T}, and both for MoS$_2$ and WS$_2$, reveal an efficient emission at $\sim 1.2-1.4$ eV, see Fig.\ \ref{fig:PL}\textbf{a}. This emission corresponds to the intralayer optical emission of InSe (either free or defect-localised \cite{InSe_PL_venanzi}), but, noticeably, its efficiency is much higher than that typically found in InSe flakes. Given the large spread in intensity (one/two orders of magnitude) that characterises InSe flakes with the same thickness \cite{InSe_PL_intensity_variability}, reliable information on the intensity of the HSs can be obtained by a direct comparison within the very same InSe flake. Therefore, in Fig.\ \ref{fig:PL}, we compare the $\mu$-PL signal recorded on the HS-bubble (IN) and right outside (OUT), $i.e.$ where InSe is deposited on the bulk TMD flake and thus not in contact with the TMD-ML bubble (see sketch in panel \textbf{b}).
Here, we note that the blue-shift of the IN InSe PL peak with respect to the OUT one apparent at \emph{T}=5 K can be ascribed to an effective increase of the number of photo-generated carriers in the InSe layer. As we show next, these carriers are injected from the strained TMD ML, eventually leading to a saturation of the InSe lower-energy localised levels and hence to a shift of the PL toward the free exciton states. This is further discussed in \textcolor{purple}{Supporting Note 3}, where $\mu$-PL studies as a function of $T$ are reported.
The InSe enhancement factor ---defined as the ratio between the $\mu$-PL peak intensities detected IN and OUT ($I_\mathrm{IN}/I_\mathrm{OUT}$, where $I$ is the peak intensity)--- may show some variation with temperature, although a univocal trend cannot be found. Nevertheless, an indubitable emission enhancement between one and three orders of magnitude is systematically observed, as shown by a statistical analysis of several HS-bubbles, see Fig.\ \ref{fig:PL}\textbf{c}. It should be noticed that the enhancement factors displayed in the figure were measured under non-resonant conditions, namely by excitation with a 532-nm laser. As shown later, even larger PL intensities can be detected by pumping in resonance with the TMD excitons.

\begin{figure*}[htpb]
\centering
\includegraphics[width=0.85\textwidth]{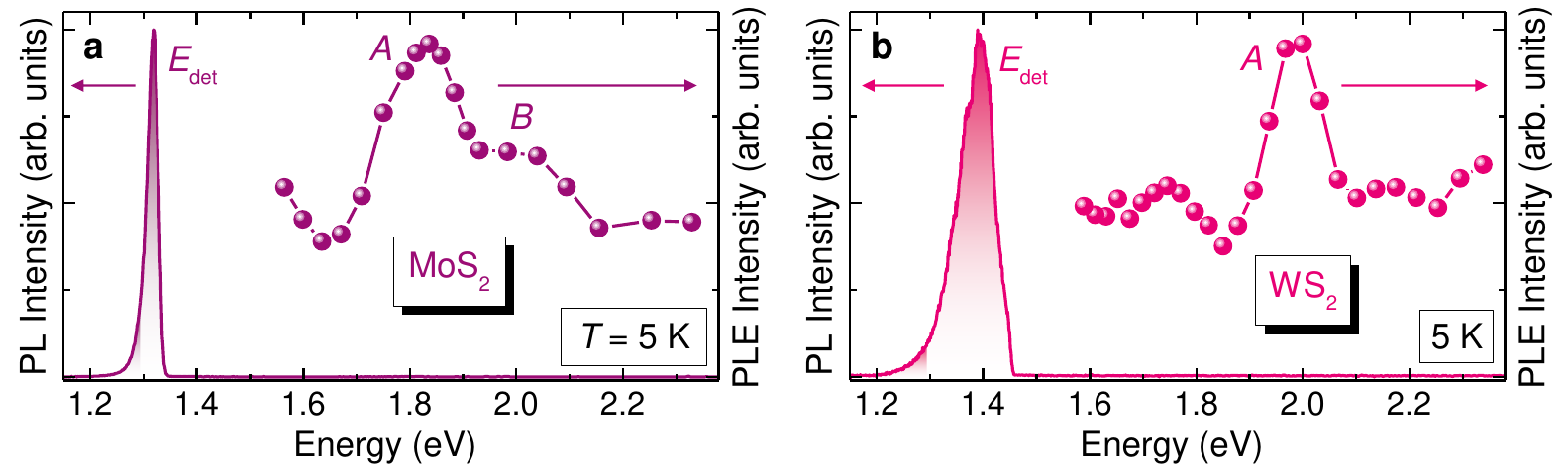}
\caption{\textbf{Energy-resonant photo-excited carrier transfer.} PLE spectra of an HS-bubble based on MoS$_2$ (\textbf{a}) and of one based on WS$_2$ (\textbf{b}). The corresponding PL band whose intensity was detected during the PLE measurements is displayed. Exciton resonances attributable to the A and B excitons of the TMD are highlighted.}
\label{fig:PLE}
\end{figure*}
As discussed in \textcolor{purple}{Supporting Note 4}, we realised several control samples to verify that there is no enhancement in the absence of strain, and that the enhancement is due to strain solely, and not to interference or exciton-dipole orientation effects.
The giant PL enhancement of InSe in InSe/MS$_2$ HS-bubbles can thus be explained by hypothesising a strain-induced charge transfer from the MS$_2$ layer toward InSe. To ascertain this, we performed $\mu$-PL excitation ($\mu$-PLE) measurements both on hBN/InSe/MoS$_2$ and on hBN/InSe/WS$_2$ HS-bubbles, see Fig.\ \ref{fig:PLE}.
Noticeably, clear exciton resonances are observed. 
For hBN/InSe/MoS$_2$ HS-bubbles, two resonances are found, at 1.84 eV and 2.01 eV. The former --attributed to the direct $A$ exciton-- is redshifted by about 0.1 eV with respect to the $A$ exciton in planar MoS$_2$ MLs due to strain \cite{blundo_review,Blundo_prr}. 
The second resonance at 2.01 eV is $\sim 0.17$ eV above the lowest energy one. Such a  distance is compatible with the $A$-$B$ exciton distance \cite{TMD_reflectance-ML_frisenda}, and we thus ascribe the 2.01 eV resonance to the $B$ exciton. In hBN/InSe/WS$_2$ HS-bubbles, instead, only one clear resonance --attributed to the $A$ exciton-- is observed at 1.98 eV, $i.e$, about 0.1 eV below the $A$ exciton in unstrained WS$_2$ MLs (similarly to the MoS$_2$ case).
$\mu$-PLE measurements were performed also in a hBN/InSe/MoS$_2$ unstrained HS: As shown in \textcolor{purple}{Supporting Note 5}, no resonances were found in the absence of strain.
The $\mu$-PLE measurements clearly point to a charge transfer from the MS$_2$ ML to the InSe flake, which is activated by the strain selectively applied to the MS$_2$ ML. As a matter of fact, a total strain of about $0.5 \%$ applied to a MoS$_2$/InSe HS as a whole resulted in no PL enhancement in Ref.\ \cite{funneling_MoS2InSe}. 
Therefore, our results clearly show that relatively high strains, applied to the MS$_2$ ML, are required to activate a charge transfer. It should also be noticed that such charge transfer is much more efficient than that previously obtained in type-I 2D HSs (such as in MoTe$_2$/WSe$_2$ HSs, where only small PL enhancement factors $\le 2$ could be obtained \cite{HSs_type_I_MoTe2_WSe2_yamaoka}), which can be attributed to a near-ideal condition achieved in our case by selective strain engineering.

To further elucidate on the effect of strain on such HSs, we performed DFT calculations.
The primary factor for ultrafast charge transfer in vdW HSs stems from the band alignment, coupled with the tendency of photoexcited electrons and holes to relax towards the CBM and VBM of the HS, respectively. In case of type-I band alignment, where both CBM and VBM reside within the same material, like in the present system (see next), upon photo-excitation both electrons and holes migrate across the layers during relaxation. DFT simulations play a pivotal role in understanding charge transfer processes in such systems, providing valuable insights into the fundamental physics and guiding material design and optimisation.

\begin{figure*}[htpb]
\centering
\includegraphics[width=0.8\textwidth]{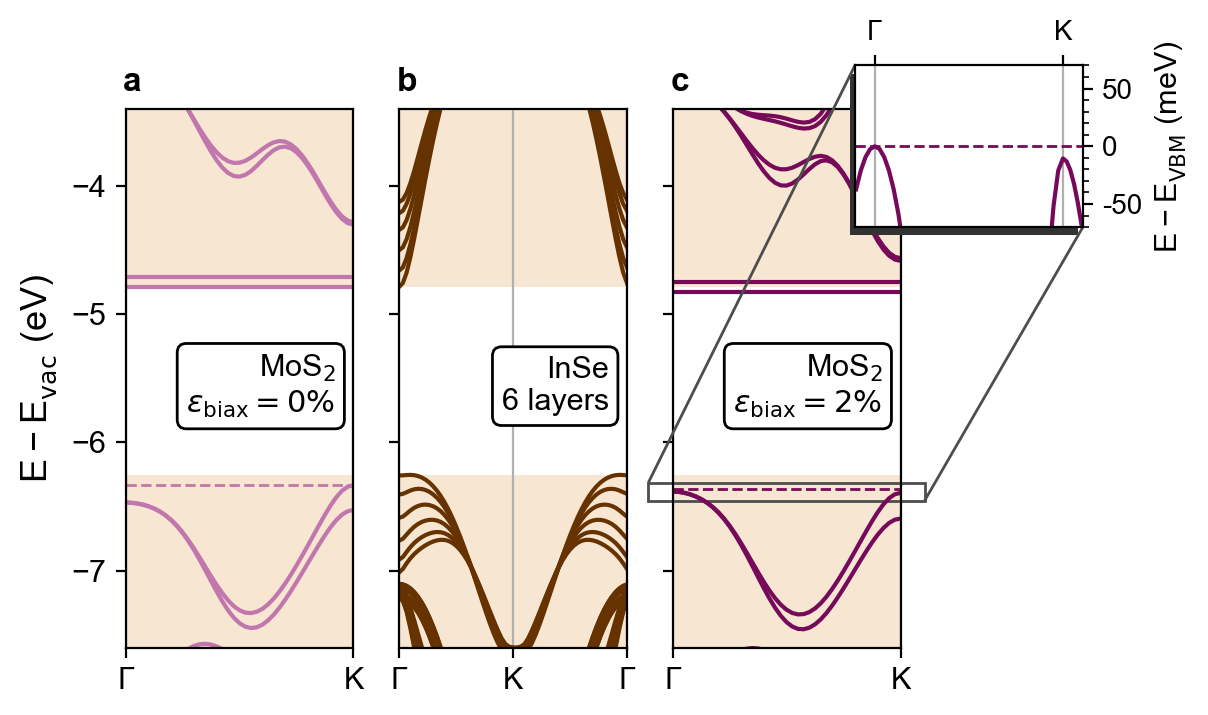}
\caption{\textbf{Heterostructured bubble band alignment.} DFT-calculated band structures of: \textbf{a} MoS$_2$ ML with S-vacancies at 0\% strain, \textbf{b} 6L InSe slab and \textbf{c} MoS$_2$ ML at 2\% biaxial strain. The band alignments with respect to the vacuum level are determined following the methodology outlined in Ref.~\cite{Kresse}, while the paths in the Brillouin zone were chosen to enhance comparability across panels. The shaded regions mark the VB and CB of 6L-InSe in all panels. The inset in panel \textbf{c} shows a zoom of the top VB of strained MoS$_2$ at the $\Gamma$ and K points.}
\label{fig:DFT}
\end{figure*}

Focusing on MoS$_2$/InSe HSs, we investigated the electronic band alignment between the ML and a 6L-InSe slab. Since defect states, and especially sulphur vacancies, play an important role in determining the electronic properties of TMDs \cite{Defects_MoS2_hong}, we also took them into account by including a sulphur vacancy in the MoS$_2$ ML in the simulation cell. The MoS$_2$ and InSe bands were aligned to the vacuum level accounting for the interface dipole following the approach of Refs.~\cite{Tanaka,Kresse}. 
The band structures were computed with the Heyd–Scuseria–Ernzerhof range-separated hybrid functional HSE06 \cite{HSE06}. 
See \textcolor{purple}{Supporting Information, Methods} for further details.
Figs.~\ref{fig:DFT}\textbf{a} and \textbf{c} show the band structure of MoS$_2$ ML at 0\% and 2\% biaxial strain, respectively. 
Such value was chosen to match the strain estimated through Raman experiments in \textcolor{purple}{Supporting Note 2}.
In agreement with previous calculations \cite{Molina-Sanchez,Strain_th_ML_TMDs_Zollner,Strain_th_ML_TMDs_johari,Strain_th_ML_MoS2-MoSe2-WS2-WSe2_chang,Strain_th_ML_WS2-MoS2_shi,Strain_th_ML_WS2-MoS2_ghorbani}, the application of 2\% tensile biaxial strain shifts the VBM from K to $\Gamma$, see inset in Fig.~\ref{fig:DFT}\textbf{c}. In the CB, the minimum at the K point rapidly shifts down in energy. On the other hand, the sulphur vacancy introduces a pair of flat states within the bandgap, which are minimally affected by strain.
The band structure of 6L-InSe is displayed in Fig.~\ref{fig:DFT}\textbf{b} to be readily compared with the MoS$_2$ unstrained and strained cases.
The computed band structures provide the electronic scenario accounting for the giant PL enhancement of InSe experimentally observed in the HS-bubbles. 
A type-I band alignment is found both in the absence and in the presence of strain, and the TMD and InSe VBs feature a sizeable overlap around the $\Gamma$ point (see Fig. \ref{fig:DFT}).
Light is strongly absorbed by the A exciton of MoS$_2$, mainly due to photo-excited electrons and holes at the K point in the Brillouin zone (BZ). When the TMD ML is unstrained, electrons and holes recombine radiatively at the K point on the picosecond scale \cite{ExcDynamics} and no carrier transfer toward InSe occurs. In contrast, the VBM upshift introduced by tensile strain allows the fast phonon-mediated relaxation of holes from K to $\Gamma$, that takes place on the femtosecond scale \cite{CarrierRelax}. In turn, the \emph{k}-space indirect character of the exciton in strained MS$_2$ slows down considerably its recombination rate (up to few nanoseconds \cite{Blundo_prr}). Then, holes in the strained MoS$_2$ ML can tunnel to InSe at those points in the BZ around $\Gamma$ where the bands of the two materials cross and tunneling can efficiently occur without energy or momentum exchange \cite{Krause2021,Rubio2023}. Photo-excited electrons in MoS$_2$ contribute to populate the spatially-localised defect states introduced by S-vacancies. As shown in Fig.~\ref{fig:DFT}, these states overlap in energy with the InSe CBM and, thanks to their \emph{k}-space delocalisation, allow for the efficient electron tunneling to InSe without momentum transfer.
The band alignment calculated at the HSE level, in Fig.~\ref{fig:DFT}, is confirmed by the results obtained from the explicit simulation of the full ML-MoS$_2$/6L-InSe interface, with the Generalised Gradient Approximation (GGA) functional, reported in \textcolor{purple}{Supporting Note 6}.
Hence, the concurrent injection of electrons and holes from the strained TMD ML (where efficient light absorption takes place) can populate the band edges of InSe and shift the PL mechanism to the more efficient free-exciton radiative recombination.
We highlight that, for increasing number of InSe layers, the bandgap slightly reduces (by about 0.1 eV in going from 6 to 10 layers), so that the results of the calculations performed for 6L-Inse can be generalised to $N$L-InSe with $6 \le N \le 10$ (used in the experiments). An analogous mechanism to that described for hBN/MoS$_2$/InSe HS-bubbles is expected also for WS$_2$-based HSs. In that case, we remark that while the CBM of WS$_2$ lies above the CBM of MoS$_2$ (by about 0.3 eV \cite{Optical_Spectroscopy_2D_shree}), S-vacancies in WS$_2$ monolayers give rise to deeper defect states, emitting about 0.5 eV below the neutral exciton \cite{SPEs_WS2_He_micevic}. In turn, the WS$_2$ defect band is expected to lie very close in energy to that of MoS$_2$ and thus quasi-resonant with the InSe CBM.

\section*{Conclusions}

In this work, we developed a novel paradigm to engineer the optoelectronic properties of 2D HSs, by demonstrating how layer-selective stretching can be efficaciously used to tailor the electronic properties of the system. Layer-selective strains can be generally achieved by deposition of 2D flakes on pre-stretched 2D materials. Here, we specifically investigated the coupling of 6-10 layer-thick InSe with (strained) MS$_2$-ML bubbles (M = Mo,W). Strain was shown to be responsible for a giant PL enhancement of the InSe signal between one and three orders of magnitude, both at cryogenic and room temperature. PLE measurements clearly proved that strain activates an electronic coupling between the HS constituent layers, entailing a charge transfer from the TMD to InSe. DFT calculations confirm that a type-I alignment is obtained, and highlight the possible mechanisms responsible for the PL enhancement. This can be attributed to a strain-induced K-to-$\Gamma$ VBM crossover along with the presence of S-vacancy states near the CBM in MS$_2$, leading to an efficient tunneling of holes (in the vicinity of $\Gamma$) and electrons (through momentum-delocalised defect states) at those points in the BZ where electronic states overlap in energy and momentum.
In turn, a 2D type-I HS characterised by an unprecedentedly efficient charge transfer --much larger than in previous TMD-based HSs-- is achieved, thanks to the nearly ideal band alignment triggered by selective strain engineering.
The significant enhancement of the PL efficiency of InSe achieved here, paired with its highly remarkable electronic and transport properties, dramatically improves the prospects for the exploitation of this material in a wide range of optoelectronic applications.



\section*{Author Contributions}
E.B. and A.Polimeni conceived and supervised the research. F.T., M.C. and E.B. fabricated the heterostructures. E.B., M.C., F.T., A.Patra, and M.F. performed the optical measurements and analysed the data. M.R.F. and M.P. performed the DFT calculations. G.P. and E.B. performed the AFM measurements and analysed the data. Z.K. and A.Patan\`{e} grew the InSe samples. T.T. and K.W. grew the hBN samples. E.B., M.C., M.R.F, M.P. and A.Polimeni wrote the manuscript. The results and the manuscript were approved by all the coauthors.\\

\section*{Acknowledgments}
The authors thank Riccardo Frisenda for discussions.
This project was funded within the QuantERA II Programme that has received funding from the European Union’s Horizon 2020 research and innovation programme under Grant Agreement No 101017733, and with funding organisations Ministero dell'Universit\'{a} e della Ricerca (A.P and M.F.) and by Consiglio Nazionale delle Ricerche (G.P.). A.P. and M.F. acknowledge financial support from the PNRR MUR project PE0000023-NQSTI. M.R.F.\ acknowledges the High-Performance Computing, Big Data, and Quantum Computing Research Centre, established under the Italian National Recovery and Resilience Plan (PNRR). 
M.F. and G.P. acknowledge funding from the PRIN2022 project DELIGHT2D (Prot. 20222HNMYE). M.P. acknowledges Union—NextGenerationEU under the Italian National Center 1 on HPC—Spoke
6: “Multiscale Modelling and Engineering Applications”.
M.R.F.\ and M.P.\ acknowledge CINECA for the availability of high-performance computing resources under the ISCRA-B and C initiatives.
Z.R.K.\ acknowledges funding through a Nottingham Research Fellowship from the University of Nottingham.
K.W.\ and T.T.\ acknowledge support from the JSPS KAKENHI (Grant Numbers 20H00354 and 23H02052) and World Premier International Research Center Initiative (WPI), MEXT, Japan.

\onecolumn


\begin{titlepage}

\begin{center}
	{\Large SUPPORTING INFORMATION for\\\mbox{}\\
		\textbf{
		Giant light emission enhancement in strain-engineered InSe/MS$_2$ (M=Mo,W) van der Waals heterostructures
		}}
\end{center}

\vspace{1 mm}

\begin{center}
	{\large Elena Blundo,$^{1,*}$ Marzia Cuccu,$^{1}$ Federico Tuzi,$^{1}$ Michele Re Fiorentin,$^{2}$ Giorgio Pettinari,$^{3}$ Atanu Patra,$^{1}$ Salvatore Cianci,$^{1}$ Zakhar Kudrynskyi,$^{4}$ Marco Felici,$^{1}$ Takashi Taniguchi,$^{5}$ Kenji Watanabe,$^{6}$ Amalia Patan\`{e},$^{8}$ Maurizia Palummo,$^{6}$ and Antonio Polimeni$^{1,*}$}
\end{center}

\begin{center}
	\textit{\mbox{}$^1$ Physics Department, Sapienza University of Rome, Piazzale Aldo Moro 5, 00185 Rome, Italy.\\
	\mbox{}$^2$ Department of Applied Science and Technology, Politecnico di Torino, corso Duca degli Abruzzi 24, 10129 Torino, Italy\\
	\mbox{}$^3$ Institute for Photonics and Nanotechnologies, National Research Council, 00133 Rome, Italy.\\
	\mbox{}$^4$ Faculty of Engineering, University of Nottingham, Nottingham, NG7 2RD, UK.\\
	\mbox{}$^5$ Research Center for Materials Nanoarchitectonics, National Institute for Materials Science,  1-1 Namiki, Tsukuba 305-0044, Japan.\\
        \mbox{}$^6$ Research Center for Electronic and Optical Materials, National Institute for Materials Science, 1-1 Namiki, Tsukuba 305-0044, Japan.\\
        \mbox{}$^7$ School of Physics and Astronomy, University of Nottingham, Nottingham, NG7 2RD, UK.\\
	\mbox{}$^8$ INFN, Dipartimento di Fisica, Universitá di Roma Tor Vergata, Via della Ricerca Scientifica 1, 00133 Rome, Italy.\\
	}
\end{center}

\mbox{}$^*$ Corresponding authors: antonio.polimeni@uniroma1.it, elena.blundo@uniroma1.it

\vspace{5 mm}

\fancyhead[LE,RO]{\tiny\thepage}
\renewcommand{\thesubfigure}{\alph{subfigure})}

\tableofcontents

\end{titlepage}

\newpage

\renewcommand{\theequation}{\arabic{section}.\arabic{equation}}
\renewcommand{\thetable}{\arabic{section}.\arabic{table}}
\renewcommand{\thefigure}{\arabic{section}.\arabic{figure}}
\renewcommand{\thesubsection}{A.\arabic{subsection}}
\setcounter{subsection}{0}
\renewcommand{\thesubfigure}{}
\sectionmark{} 


        \section*{Methods}
        \addcontentsline{toc}{section}{
        Methods
        }

\noindent
\textbf{Sample fabrication}\\
The HSs were fabricated by the standard dry transfer technique. TMD flakes were mechanically exfoliated by the scotch tape method and deposited on SiO$_2$/Si substrates. The samples were then loaded into a Kaufman chamber and exposed to hydrogen-ion atoms with energies of 10-20 eV, according to the procedure discussed in refs.\ \cite{Tedeschi_AdvMater,blundo_hBN-bubbles_nanolett}, leading to the formation of bubbles. InSe was mechanically exfoliated by the scotch tape method and deposited on PDMS. Thin flakes were identified and deposited over some selected bubbles with a 2D transfer stage (by HQ Graphene). Thin hBN flakes were exfoliated and deposited with the same procedure.\\

\noindent
\textbf{Atomic force microscopy}\\
AFM measurements were performed using a Veeco Digital Instruments Dimension D3100 microscope equipped with a Nanoscope IIIa controller, employing Tapping Mode monolithic silicon probes with a nominal tip curvature radius of $5-\SI{10}{\nano\meter}$ and a force constant of \SI{40}{\newton\per\metre}. All the scans were performed at room temperature and at ambient conditions. All the data were analysed with the Gwyddion software. \\

\noindent
\textbf{$\mu$-PL measurements}\\
For $\mu$-PL measurements, the samples were placed in a closed-circuit He cryostat by Montana Instruments; vacuum conditions were established (chamber pressure $< 2$ mTorr) and the sample was brought to the desired temperature (or kept at RT). The excitation laser was provided by a single frequency Nd:YVO$_4$ laser (DPSS series by Lasos) emitting at 532 nm. To avoid oxidation effects, we employed low laser powers, typically equal to 4 $\mu$W (and always below 40 $\mu$W). The luminescence signal was spectrally dispersed by a 20.3 cm focal length Isoplane 160 monochromator (Princeton Instruments) equipped with a 150 grooves/mm and a 300 grooves/mm grating and detected by a back-illuminated N$_2$-cooled Si CCD camera (100BRX by Princeton Instruments). The laser light was filtered out by a very sharp long-pass Razor edge filter (Semrock). A 100$\times$ long-working-distance Zeiss objective with NA = 0.75 was employed to excite and collect the light, in a backscattering configuration and using a confocal setup.\\

\noindent
\textbf{$\mu$-PL excitation measurements}\\
For $\mu$-PL excitation ($\mu$-PLE), we employed the same ps supercontinuum laser used for tr $\mu$-PL. The laser wavelength was automatically changed by an acousto-optic tunable filter and employing a series of shortpass and longpass filters to remove spurious signals from the laser. The detection wavelength was selected using the same monochromator and detector employed for cw $\mu$-PL measurements.\\

\noindent
\textbf{$\mu$-Raman measurements}\\
$\mu$-Raman measurements were taken in the same experimental configuration employed for $\mu$-PL measurements. In this case, the Raman signal was spectrally dispersed by a 75 cm focal length Acton monochromator (Princeton Instruments) equipped with a 12000 grooves/mm grating.\\

\noindent
\textbf{Density functional theory calculations}\\
Simulations were performed with DFT using the plane-wave expansion method as implemented in the Quantum ESPRESSO package \cite{Giannozzi_2009,Giannozzi_2017}.

Fully relativistic, norm-conserving pseudopotentials \cite{pseudo_dojo} were employed to account for spin-orbit coupling. Convergence was achieved considering an 80~Ry kinetic energy cutoff and an $8\times8\times1$ ($9\times9\times1$) Monkhorst-Pack $k$-point mesh for the 6L-InSe slab  (TMD monolayer). The $k$-point meshes were accordingly rescaled in supercell calculations. The optimised geometries of TMD MLs and 6L-InSe were obtained by performing structure relaxations with the Perdew-Burke-Ernzerhof (PBE) functional \cite{PBE}. A 20~\AA{} vacuum region along the direction perpendicular to the slab plane was introduced to ensure decoupling of periodic replicas. Structure relaxations were assumed to have reached convergence when the maximum component of the residual ionic forces was smaller than $10^{-8}$~Ry/Bohr. 
The sulphur vacancy in the TMD monolayer was simulated considering a $5\times5$ supercell.

The Heyd–Scuseria–Ernzerhof range-separated hybrid functional (HSE06) \cite{HSE06} was chosen over PBE to calculate the electronic bandstructures and band alignments due to its superior capability in accurately describing electronic properties. While the PBE functional is commonly employed for structure optimisations owing to its computational efficiency, it frequently produces underestimated bandgaps, which can lead to incorrect predictions of band alignments. The HSE06 functional represents a reasonable and practical choice to investigate charge transfer and electronic properties, providing accurate results at a lower computational cost compared to more advanced methods like GW calculations.

The band alignment in HSs is affected by dipoles originating from the atomic configuration and orientation of the two contacting materials.
To account for the interface dipole between the TMD and InSe, we adopted the methodology outlined in Ref.~\cite{Kresse}, which adjusts the initial alignment derived from isolated phases by considering the shift in the inner potential between the two phases in a model, lattice-matched, interface. However, while advantageous for their reduced number of atoms, lattice-matched interfaces inherently introduce undesired strain in the two phases. This is then compensated for by considering the potential offset between the strained and unstrained lattices.

Additional defects in InSe, such as Se-vacancies, have not been explicitly included in the simulation since they induce localised electronic states below the CBM or above the VBM, that can be readily saturated by the injected carriers \cite{Mudd2014}.

Finally, to verify the predictions and prove the type-I band-alignment of the real interface, we simulated supercells of the explicit ML-MoS$_2$/6L-InSe systems. The interfaces between 6L-InSe and 0\%- and 2\%-strained TMD MLs were obtained with the CellMatch code \cite{CellMatch} (see \textcolor{purple}{Supporting Note 6}) and are reported in Supporting Figure 6.1. Due to the large number of electrons (3504 and 3224 electrons for the unstrained and strained HS, respectively), the geometry optimisation and the analysis of the electronic properties were performed with the rev-vdW-DF2 \cite{vdW-DF2} GGA functional, without spin-orbit coupling. 

\newpage
	
	\section*{
        Supporting Note 1. Heterostructured bubble morphology \emph{vs} temperature
        }
	\addcontentsline{toc}{section}{
        Supporting Note 1. Heterostructured bubble morphology \emph{vs} temperature
        }
	\setcounter{section}{1}
        \sectionmark{}
	\setcounter{figure}{0}
	\setcounter{equation}{0}
        \setcounter{table}{0}

\begin{figure}[htbp]
        \begin{center} 
        \includegraphics[width=1.0\textwidth]{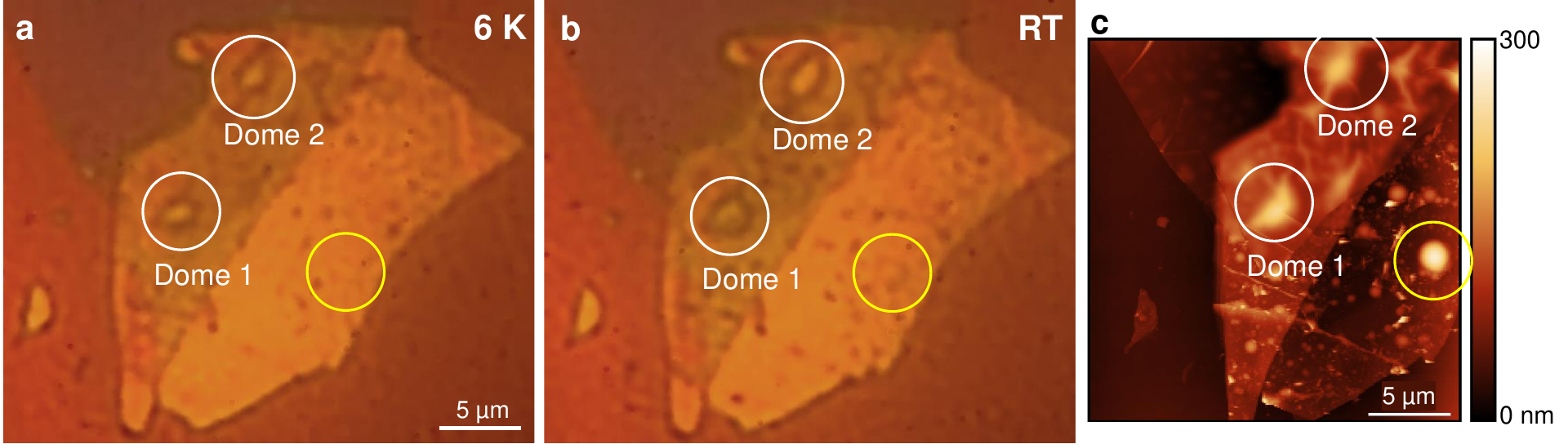}
        \caption{
        \textbf{Morphology of heterostructured bubbles \emph{vs} temperature.} \textbf{a} Optical image acquired at 6 K of two hBN/InSe/WS$_2$ HS-bubbles, highlighted by the white circles. The yellow circle highlights the position of a WS$_2$ bubble which was not covered neither by InSe nor by hBN. \textbf{b} Same as \textbf{a} but at RT. \textbf{c} AFM image of the same flake of panels \textbf{a} and \textbf{b}, acquired at RT.
        }
        \label{fig:HS_dome_vs_T_optical_AFM}
        \end{center}
        \end{figure}

        The heterostructuring process employed in this work ---consisting in the deposition of thin InSe and hBN flakes on transition metal dichalcogenide (TMD) bubbles--- has profound implications on the mechanics of the system in the low temperature regime. In fact, hydrogen-filled TMD bubbles are known to suddenly deflate at about 30 K due to the gas-to-liquid phase transition of H$_2$ \cite{Tedeschi_AdvMater,Cianci_SPEs_WS2_domes}, while we do not observe any major change in the morphology of our heterostructured bubbles (HS-bubbles) even at 5 K. 
        This is shown, as an example, for two hBN/InSe/WS$_2$ HS-bubbles in Fig.\ \ref{fig:HS_dome_vs_T_optical_AFM}: Figs.\ \ref{fig:HS_dome_vs_T_optical_AFM}\textbf{a} and \textbf{b} show the optical images of a flake containing the two HS-bubbles, taken at 6 K and room-temperature (RT), respectively. The two HS-bubbles are highlighted by white circles. The yellow circle, instead, highlights the position of a WS$_2$ bubble which was not covered neither by InSe nor by hBN. Indeed, the two HS-bubbles are clearly visible in both images, while the non-capped bubble is not visible in the image acquired at 6 K since it deflated at around 30 K with the occurring of the gas-to-liquid transition of the hydrogen inside the bubble. The same bubble can in fact be seen at RT, as it can be noticed by looking at the optical image shown in panel \textbf{b}. Finally, Fig.\ \ref{fig:HS_dome_vs_T_optical_AFM}\textbf{c} shows an atomic force microscope (AFM) image of the same flake, at RT.

        Noticeably, the two HS-bubbles show only a moderate size variation between RT and 6 K. 
        We attribute this effect to the role played by the InSe and hBN flakes, that adapt to the bubble underneath during the deposition process, minimising the total energy of the system. The deflation of the bubble when H$_2$ liquefies would, in turn, imply a mechanical deformation of the above flakes and thus an energy cost. The cost of keeping the bubble in-shape is likely lower, leading to this peculiar effect for which the strain is maintained even below 30 K, when no internal pressure is exerted by hydrogen on the TMD ML. Raman studies, discussed in \textcolor{purple}{Supporting Note 2}, corroborate this finding by demonstrating that only a moderate strain reduction is observed when decreasing $T$ from RT to 5 K.
        
        This mechanical effect is similar to that played by tied arches in bridges. In tied-arch bridges (HS-bubbles), the main beam (TMD dome) is coupled to the arch (InSe+hBN) via suspenders (van der Waals forces). The rigidity of the arch (InSe+hBN) and strength of the suspenders (van der Waals adhesion) make it so that the whole structure maintains its shape even when the main beam (TMD dome) is subject to external forces, $e.g.$ the gravity and when heavy vehicles pass through the bridge (the hydrogen gas undergoes a gas-to-liquid transition).

        The mechanical effect observed of our micro-domes represented by the HS-bubbles also resembles that of the giant (external diameter = 54.8 m) and beautiful Brunelleschi's Dome in Florence. Indeed, that Dome is characterised by a double-layer structure, which makes the Dome structure robust, similarly to our HS-micro-domes.

    \clearpage
    \newpage
    
    \section*{Supporting Note 2. Strain characterisation in heterostructured bubbles}
	\addcontentsline{toc}{section}{Supporting Note 2. Strain characterisation in heterostructured bubbles}
	\setcounter{section}{2}
        \sectionmark{}
	\setcounter{figure}{0}
	\setcounter{equation}{0}
        \setcounter{table}{0}

\begin{figure*}[htbp]
        \begin{center} \includegraphics[width=0.75\textwidth]{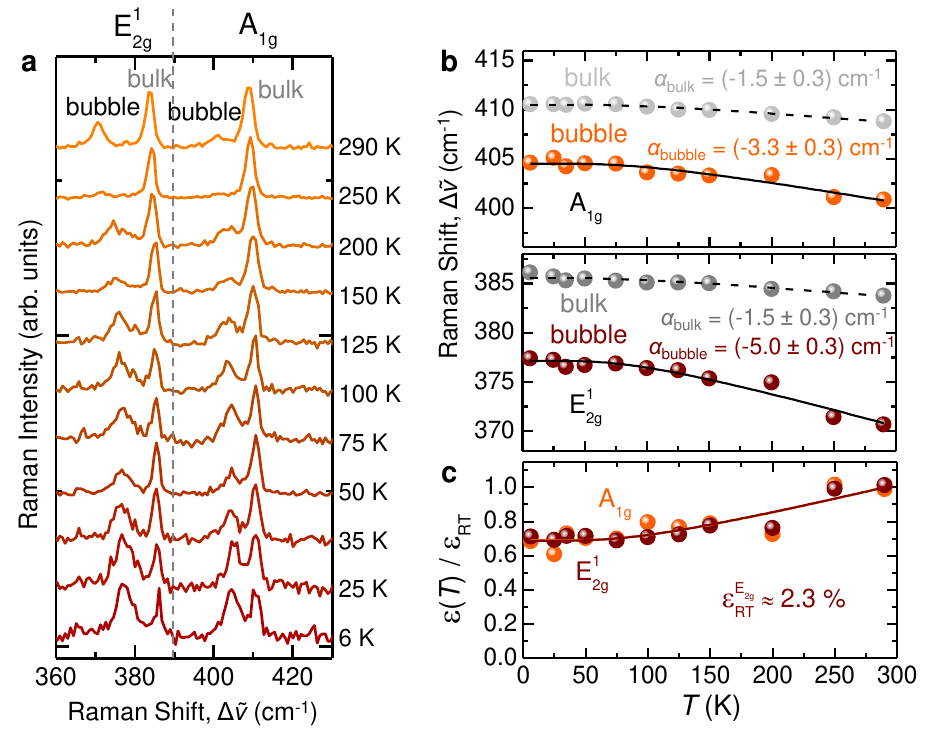}
        \caption{
        \textbf{Structural properties of a heterostructured bubble \emph{vs} temperature investigated by Raman spectroscopy.} \textbf{a} Raman spectra of a hBN/InSe/MoS$_2$ HS-bubble acquired in the phonon regions of the MoS$_2$ in-plane $E^1_\mathrm{2g}$ and of the out-of-plane $A_\mathrm{1g}$ modes at different temperatures, stacked by y-offset. The Raman modes of both the bubble and the bulk beneath the bubble can be seen. \textbf{b} Raman shifts of the $A_\mathrm{1g}$ and $E^1_\mathrm{2g}$ modes of both the bubble and the bulk $vs$ $T$. The solid (dashed) lines are fits to the data concerning the bubble (bulk) via Eq.\ \ref{eq:Klemens}. The $\alpha$ constants obtained by fitting the data are displayed.
        \textbf{c} Strain variation with temperature calculated starting from the experimental Raman shifts shown in panel \textbf{b} with respect to the room-temperature (RT, corresponding to 290 K) value, estimated following Eq.\ \ref{eq:strain_Raman} as:        ${\varepsilon(T)/\varepsilon_\mathrm{RT}} = [ {\Delta \tilde{\nu}}^\mathrm{ML} ( T ) - {\Delta \tilde{\nu}} ^\mathrm{bubble} ( T )] / [ {\Delta \tilde{\nu}}^\mathrm{ML} ( \mathrm{RT} ) - {\Delta \tilde{\nu}} ^\mathrm{bubble} ( \mathrm{RT} )]$.
        The solid lines display the strain calculated with the same approach, but starting from the fitting curves displayed in panel \textbf{b}.
        }
        \label{fig:Raman}
        \end{center}
        \end{figure*}

    We investigated the structural properties of the HS-bubbles by performing Raman measurements as a function of $T$ on hBN/InSe/MoS$_2$ bubbles. Specifically, we monitored the Raman signal of MoS$_2$ in the phonon regions of the in-plane $E^1_\mathrm{2g}$ and of the out-of-plane $A_\mathrm{1g}$ modes.
    Fig.\ \ref{fig:Raman}\textbf{a} shows the micro-Raman spectra acquired on a hBN/InSe/MoS$_2$ bubble as a function of temperature. The Raman spectra show the appearance of both the $E^1_\mathrm{2g}$ and $A_\mathrm{1g}$ modes of the bubble and those of the bulk flake beneath the bubble (with the bubble peak being at lower frequency than the bulk one, for both the modes). The bulk peaks feature a moderate thermal shift with temperature. The bubble peaks feature a larger shift, suggesting a strain variation, as testified by the quantitative analysis displayed in panel \textbf{b}, where the Raman shift values ($\Delta \tilde{\nu}$) of all modes are shown.

    The Raman shift values were fitted with the formula
    \begin{equation}
    \Delta \tilde{\nu} \left( T \right) = \Delta \tilde{\nu}_0 + \alpha \left( 1+ \frac{2}{e^{\frac{h c \Delta \tilde{\nu}_0}{2 k_{\mathrm{B}}T}}-1} \right)
    \label{eq:Klemens}
    \end{equation}
    where $\Delta \tilde{\nu}_0$ and $A$ are constants (left as fitting parameters) and only the lower-order three-phonon processes of the anharmonic Klemens model are taken into account via the parameter $\alpha$ \cite{klemens1966anharmonic, balkanski1983anharmonic,Cianci_SPEs_WS2_domes} ($h$ is the Planck constant, $c$ is the speed of light, and $k_\mathrm{B}$ is the Boltzmann constant).

    The Raman mode frequency of the reference unstrained monolayer (ML), ${\Delta \tilde{\nu}}^\mathrm{{ML}} \left( T \right)$, can be obtained from the measured bulk mode by just adding (subtracting) a small 1.5 cm$^{-1}$ rigid shift to the E$_{\mathrm{2g}}$ (A$_{\mathrm{1g}}$) peak \cite{blundo_adhesion_prl}. Therefore, from the frequency position of the Raman peaks associated to the bulk and to the bubble modes, information on the biaxial strain $\varepsilon_\mathrm{biax}$ at different temperatures can be extracted quantitatively by applying the following formula:
    \begin{equation}
        {\varepsilon_\mathrm{biax}(T)} = \frac{1}{2} \left[ \frac{ {\Delta \tilde{\nu}}^\mathrm{ML} \left( T \right) - {\Delta \tilde{\nu}} ^\mathrm{bubble} \left( T \right)}{ \frac{d{\Delta \tilde{\nu}}}{d \varepsilon_\mathrm{tot}} } \right]
        \label{eq:strain_Raman}
    \end{equation}
    where 
    $\frac{d{\Delta \tilde{\nu}}}{d \varepsilon_\mathrm{tot}}$ is the shift rate of the Raman mode with (total) strain 
    (assumed to be constant at all temperatures \cite{postmus1968pressure})
    and the total strain is divided by 2 to obtain the biaxial strain, $i.e.$ the strain induced along just one direction (\emph{i.e.}, the radial or the circumferential one, which are equivalent at the top of the bubble \cite{blundo_adhesion_prl}).
    Fig.\ \ref{fig:Raman}\textbf{c} shows the strain variation with temperature, calculated as the strain at temperature $T$ normalised to the strain at room-temperature (RT). 
    A quantitative estimate of the in-plane biaxial strain can be obtained by the knowledge of the shift rate of the in-plane Raman mode, which is equal to ($3.24 \pm 0.41$) cm$^{-1} / \%$ \cite{blundo_adhesion_prl}.
    In turn, a biaxial strain of about $2.3 \%$ is estimated at RT, with $\sim 30 \%$ decrease at cryogenic temperatures ($i.e.$, $\varepsilon_\mathrm{biax} \approx 1.6 \%$).
    \begin{figure*}[htbp]
        \begin{center} \includegraphics[width=0.75\textwidth]{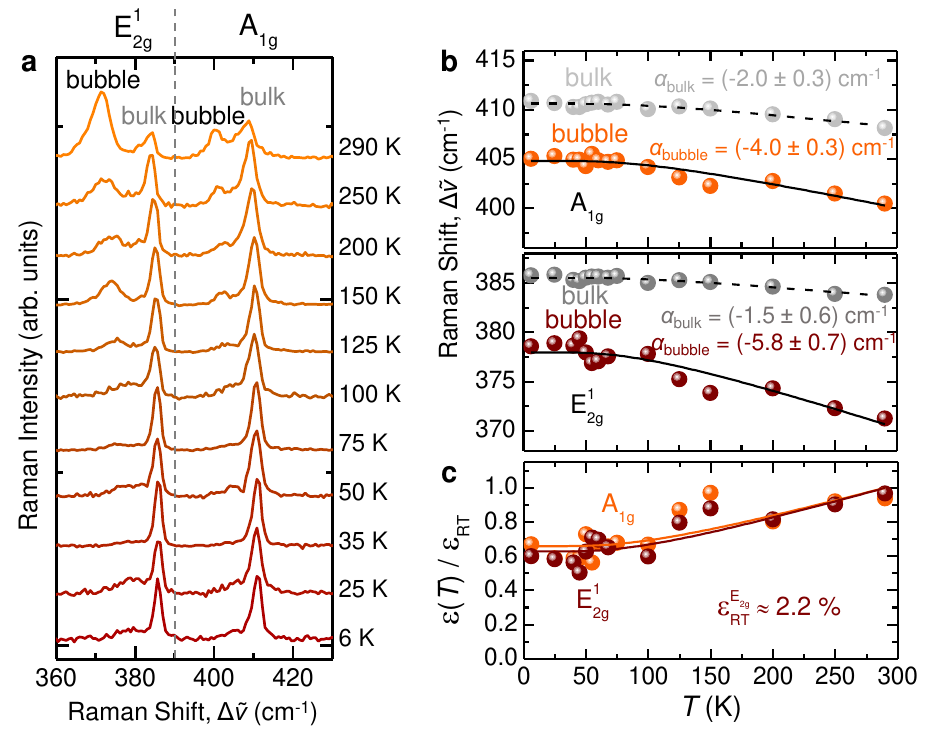}
        \caption{
        \textbf{Structural properties of another heterostructured bubble \emph{vs} temperature investigated by Raman spectroscopy.} \textbf{a} Raman spectra of another hBN/InSe/MoS$_2$ HS-bubble acquired in the phonon regions of the MoS$_2$ in-plane $E^1_\mathrm{2g}$ and of the out-of-plane $A_\mathrm{1g}$ modes at different temperatures, stacked by y-offset. The Raman modes of both the bubble and the bulk beneath the bubble can be seen. \textbf{b} Raman shifts of the $A_\mathrm{1g}$ and $E^1_\mathrm{2g}$ modes of both the bubble and the bulk $vs$ $T$. The solid (dashed) lines are fits to the data concerning the bubble (bulk) via Eq.\ \ref{eq:Klemens}. The $\alpha$ constants obtained by fitting the data are displayed.
        \textbf{c} Strain variation with temperature calculated starting from the experimental Raman shifts shown in panel \textbf{b} with respect to the room-temperature (RT, corresponding to 290 K) value, estimated following Eq.\ \ref{eq:strain_Raman} as:        ${\varepsilon(T)/\varepsilon_\mathrm{RT}} = [ {\Delta \tilde{\nu}}^\mathrm{ML} ( T ) - {\Delta \tilde{\nu}} ^\mathrm{bubble} ( T )] / [ {\Delta \tilde{\nu}}^\mathrm{ML} ( \mathrm{RT} ) - {\Delta \tilde{\nu}} ^\mathrm{bubble} ( \mathrm{RT} )]$.
        The solid lines display the strain calculated with the same approach, but starting from the fitting curves displayed in panel \textbf{b}.
        }
        \label{fig:Raman_2}
        \end{center}
        \end{figure*}

        We repeated a similar analysis for a second hBN/InSe/MoS$_2$ bubble, as shown in Fig.\ \ref{fig:Raman_2}. The results found in this second case are consistent with those found for the previous bubble. In this case, a RT biaxial strain equal to about $2.2 \%$ is estimated, with a $\sim 30-40 \%$ decrease at cryogenic temperatures ($i.e.$, $\varepsilon_\mathrm{biax} \approx 1.3-1.5 \%$).

        \newpage
	\section*{Supporting Note 3. Photoluminescence studies of heterostructured bubbles \emph{vs} temperature}
	\addcontentsline{toc}{section}{Supporting Note 3. Photoluminescence studies of heterostructured bubbles \emph{vs} temperature}
	\setcounter{section}{3}
        \sectionmark{}
	\setcounter{figure}{0}
	\setcounter{equation}{0}
        \setcounter{table}{0}

\begin{figure*}[htbp]
        \begin{center} \includegraphics[width=0.40\textwidth]{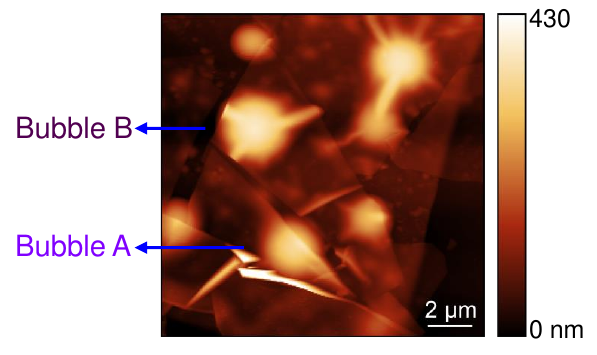}
        \caption{
        \textbf{Atomic force microscope image of two heterostructured bubbles studied by photoluminescence measurements \emph{vs} temperature.} RT AFM image of the two hBN/InSe/MoS$_2$ HS-bubbles investigated by PL measurements as a function of $T$.
        }
        \label{fig:PL_vs_T_MoS2_AFM}
        \end{center}
        \end{figure*}

\begin{figure*}[htbp]
        \begin{center} \includegraphics[width=0.95\textwidth]{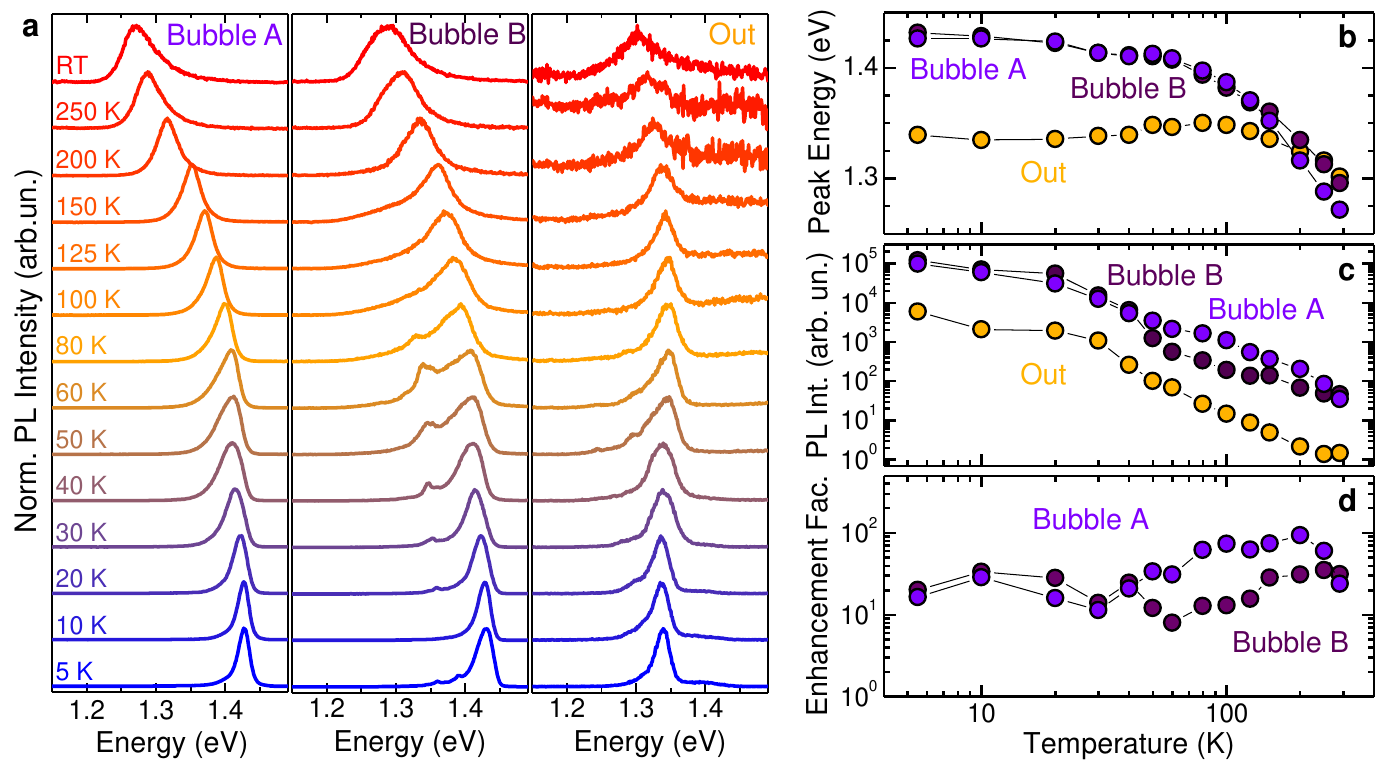}
        \caption{
        \textbf{Photoluminescence emission of heterostructured bubbles \emph{vs} temperature.} \textbf{a} PL spectra of two hBN/InSe/MoS$_2$ HS-bubbles (whose AFM image is shown in Fig.\ \ref{fig:PL_vs_T_MoS2_AFM}) and PL signal acquired on a flat region (Out) at different temperatures, in the region of the InSe signal, stacked by y-offset. \textbf{b} Peak energy of the InSe band for the two HS-bubbles and the Out region, as a function of $T$. \textbf{c} PL peak intensity of signal from the two HS-bubbles and from the Out region. \textbf{d} PL enhancement factor calculated for the two HS-bubbles from the intensity values of panel \textbf{c}.
        }
        \label{fig:PL_vs_T_MoS2_1}
        \end{center}
        \end{figure*}

        To verify how the PL enhancement observed in our selectively strained HSs evolves with temperature, we performed PL measurements as a function of $T$ on two hBN/InSe/MoS$_2$ HS-bubbles. The AFM image of the selected HS-bubbles is shown in Fig.\ \ref{fig:PL_vs_T_MoS2_AFM}. The wrinkled structure on top of the bubbles is due to the hBN capping. 
        Fig.\ \ref{fig:PL_vs_T_MoS2_1}\textbf{a} shows the result of $\mu$-PL measurements performed on the two HS-bubbles, in the energy region of the InSe PL emission. For comparison, the PL measurements acquired on a flat region of the sample (Out), where InSe lies over the MoS$_2$ bulk flake, are also displayed. A sizable shift of the PL band with $T$ is evident for both HS-bubbles, as also highlighted by the quantitative analysis of the peak energy presented in panel \textbf{b}. To the contrary, the PL band in the Out region undergoes minor changes in energy, with the peak energy displaying an S-shape behaviour when decreasing $T$. Such an S-shape behaviour was already reported in the literature for few-layer-thick InSe, and was attributed to the dominant presence of defect-related recombination \cite{InSe_PL_venanzi}. Noticeably, a rather different behaviour is observed for the HS-bubbles with respect to the Out region. Such behaviour resembles that typical of bandgap-related recombination, and can be explained by hypothesising that the strain-activated charge carrier tunnelling from the TMD to InSe favours the creation and recombination of electron-hole pairs from the band edges of InSe, rather than from defect states.
        As for the PL intensity, Fig.\ \ref{fig:PL_vs_T_MoS2_1}\textbf{c}, top panel, shows instead the PL peak intensity registered for the two bubbles and for the flat region outside, as a function of temperature. Indeed, in all cases, the PL signal intensity increases with decreasing $T$. The PL signal collected from the HS bubbles is more intense (by one-to-two orders of magnitude) than the signal acquired in the Out region all over the temperature range, as  also highlighted in panel \textbf{d}, where the enhancement factor for the two HS-bubbles is displayed. No clear trend with temperature is observed, while the enhancement shows some oscillations with $T$. This can be attributed to the combined effect of the bandgap variation of both InSe and MoS$_2$ with $T$, and of some strain variations occurring with $T$, likely resulting in a better electronic coupling between the constituent materials of the HS for some specific temperature values.

        %

        %
        %

        \clearpage


	\newpage
	\section*{Supporting Note 4. Photoluminescence studies of several control samples}
	\addcontentsline{toc}{section}{Supporting Note 4. Photoluminescence studies of several control samples}
	\setcounter{section}{4}
        \sectionmark{}
	\setcounter{figure}{0}
	\setcounter{equation}{0}
        \setcounter{table}{0}

        To verify that the emission enhancement observed in our HS-bubbles is due to the effect of strain solely, and not (i) to the fact that the InSe layer is coupled to a MS$_2$ ML, (ii) to a Fabry-Per\'{o}t effect caused by the presence of a transparent H$_2$-filled bubble underneath, or (iii) to exciton-dipole orientation effects related to the curved nature of our HSs, we prepared the following control samples: (i) planar (unstrained) hBN/InSe/MS$_2$-ML HSs deposited atop of a bulk hBN flake;
        (ii)-(iii) HS-bubbles where the bubble is made of hBN instead of MS$_2$ (the hBN bubble was still created by hydrogen-ion irradiation \cite{blundo_hBN-bubbles_nanolett}). 
\begin{figure*}[htpb]
\begin{center}
\includegraphics[width=0.75\textwidth]{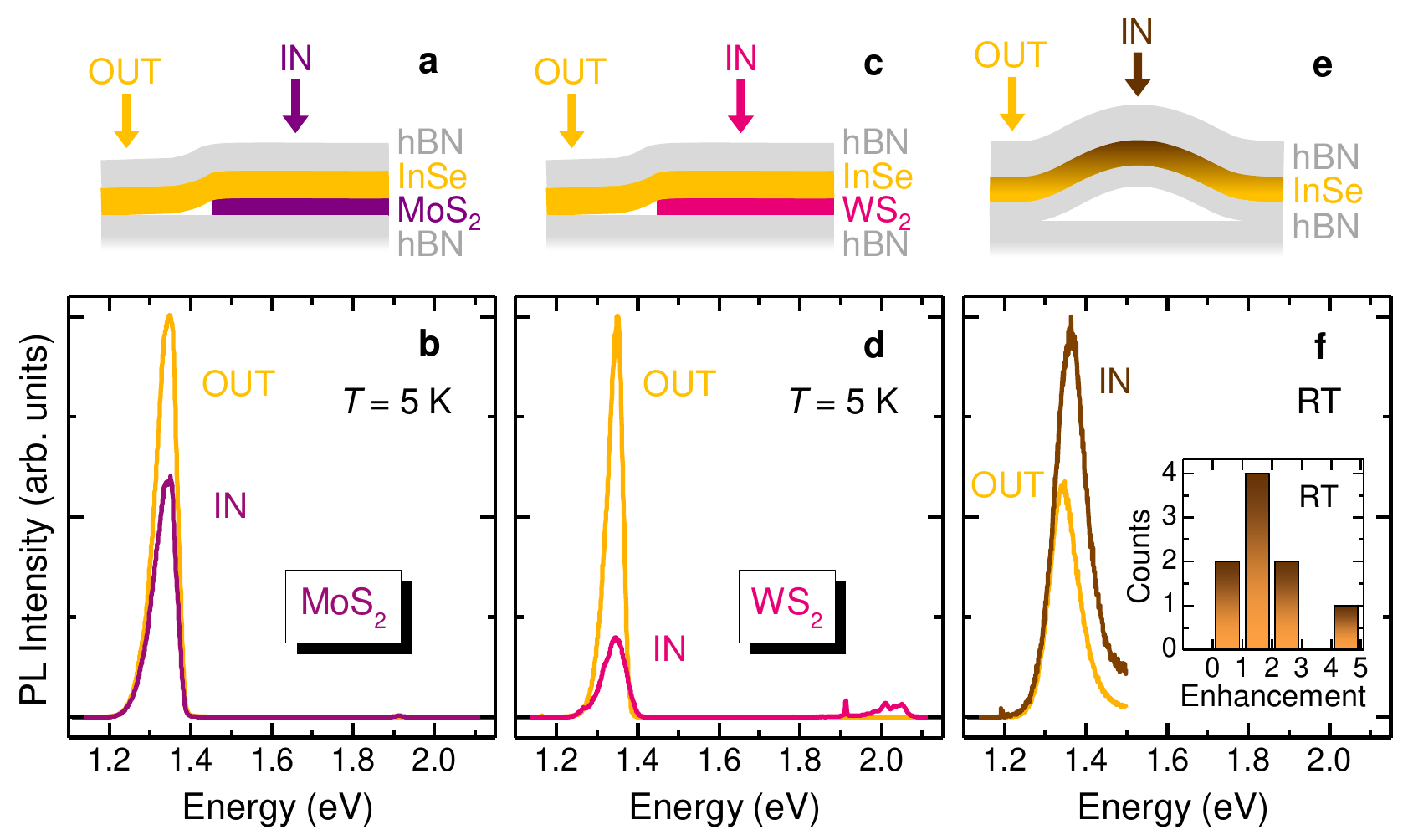}
\caption{ \textbf{Photoluminescence measurements of control heterostructures}. \textbf{a}-\textbf{d} Sketches and PL spectra of planar (unstrained) hBN/InSe/MS$_2$ HSs (MoS$_2$ in panels \textbf{a}-\textbf{b}, WS$_2$ in panels \textbf{c}-\textbf{d}). A comparison with the PL spectra in the regions where the TMD is not present and thus only InSe is present is provided. \textbf{e}-\textbf{f} Sketch (\textbf{e}) and PL spectrum (\textbf{f}) of an hBN/InSe/hBN HS-bubble ($i.e.$, where the TMD bubble is replaced by a hBN bubble). The PL spectrum acquired on the bubble is compared with the PL spectrum right outside the bubble. Inset: Statistical analysis of the enhancement factor ($I_\mathrm{IN}/I_\mathrm{OUT}$) measured in several hBN/InSe/hBN HS-bubbles at RT, with 8L- and 9L-thick InSe flakes. For the samples of panels \textbf{b}, \textbf{d} and \textbf{f} a 8L-thick InSe was used. }
\label{fig:check}
\end{center}
\end{figure*}
    The $\mu$-PL spectra of the control samples are shown in Fig.\ \ref{fig:check}. For case (i), Figs.\ \ref{fig:check}\textbf{a}-\textbf{d} show HSs formed by depositing a 8L-InSe sample on a planar (\emph{i.e.}, strain-free) MoS$_2$ (\textbf{a}-\textbf{b}) or WS$_2$ (\textbf{c}-\textbf{d}) ML. A decrease of the InSe PL signal is observed in both cases (like in Ref.\ \cite{funneling_MoS2InSe}), contrary to the PL enhancement observed in the strained case. Analogous PL reductions are also found in similar HSs with different InSe thickness, as shown later. For cases (ii)-(iii), Figs.\ \ref{fig:check}\textbf{e}-\textbf{f} show the instance of a 8L-InSe on a hBN bubble, where, instead, a PL increase by about a factor 2 is found. Similar measurements were performed on other hBN/InSe/hBN HS-bubbles, and $I_\mathrm{IN}/I_\mathrm{OUT}$ values between 0.3 and 4.4 were observed, see inset of Fig.\ \ref{fig:check}\textbf{f}. These moderate signal reductions or --more frequently-- enhancements can be likely attributed to interference effects and/or to the effect of  dipole selection rules related to light polarisation. Considering the second case, we note that the bubbles have a softly curved structure with a height-to-radius ratio of $\approx 0.11$ for hBN \cite{blundo_hBN-bubbles_nanolett} and $\approx 0.165$ for MoS$_2$ and WS$_2$ \cite{blundo_adhesion_prl} (corresponding to an inclination of at most 13° and 20°, respectively, at the very edge of the bubbles), while the laser is polarised parallel to the substrate plane. Given the out-of-plane dipole of InSe \cite{InSe_high_mu_nat_nan,InSe_dir_indir,InSe_dipole_brotons-gisbert}, this can lead to an increased light absorption in the inclined InSe layer with respect to the planar situation. PL enhancements of a factor of 2-3 were indeed proved for multilayer ($N >3$) InSe flakes bent on pillars with an inclination up to 9° \cite{InSe_pillars_mazumder} and for thick InSe flakes in the shape of ridges with a 10°-30° deflection \cite{Shi_InSe_ridge}. Our control measurements on hBN bubbles (Fig.\ \ref{fig:check}\textbf{f}) show consistent results. Such enhancements are indeed much smaller than those found for our strained TMD-based HSs.
 
\begin{figure*}[htbp]
        \begin{center} \includegraphics[width=0.85\textwidth]{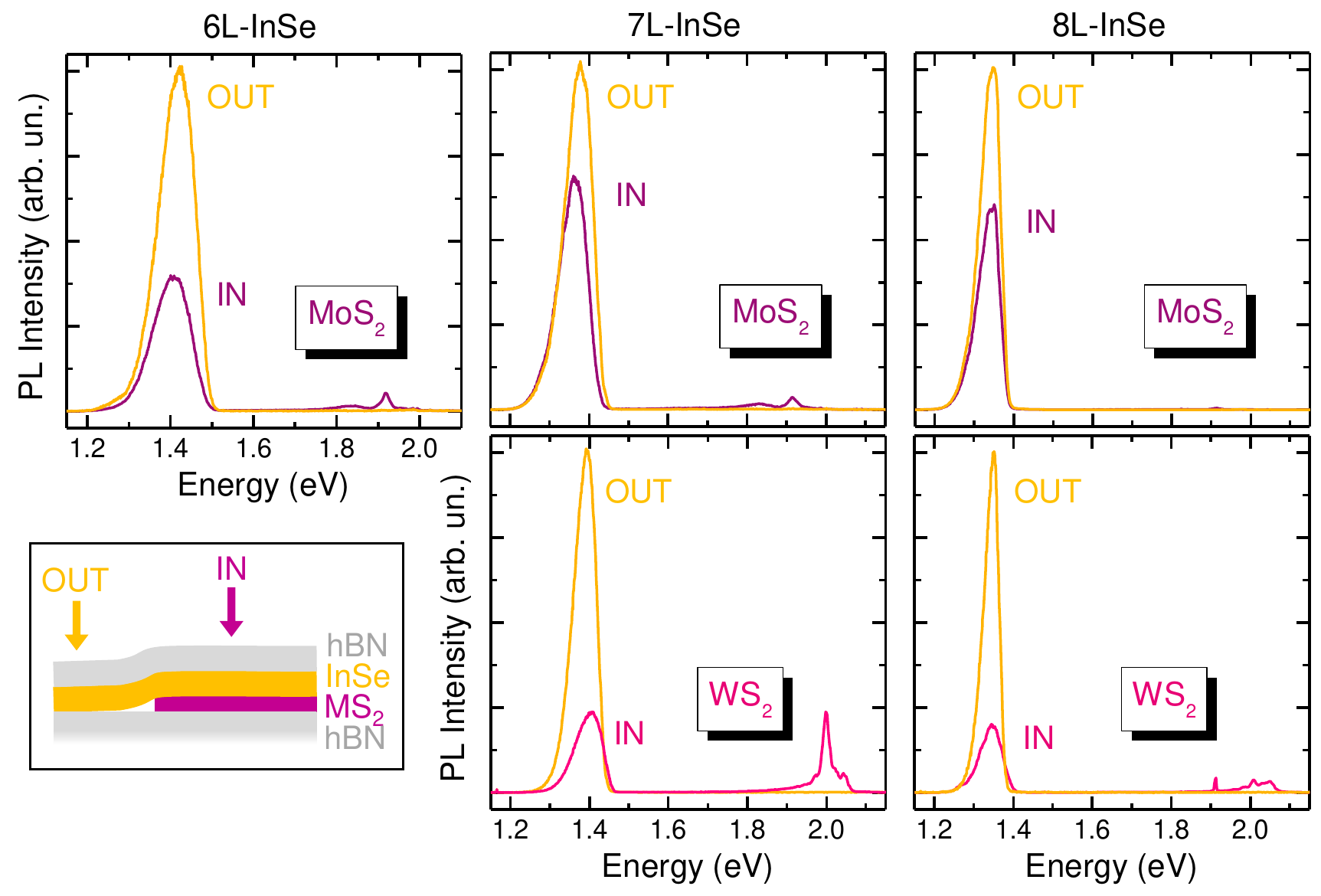}
        \caption{
        \textbf{Photoluminescence measurements of control planar heterostructures}. PL spectra of planar ($i.e.$, where the MS$_2$ ML is not strained) hBN/InSe/MS$_2$ HS (IN, see sketch at the bottom-left), and comparison with the PL spectra in the regions where the TMD is not present and thus only InSe is present (OUT, see sketch at the bottom-left). Different InSe flake thicknesses, in between 6 and 8 layers, are considered.
        }
        \label{fig:check_planar}
        \end{center}
        \end{figure*}

    Fig.\ \ref{fig:check_planar} shows the results of PL measurements performed on further unstrained HSs formed by depositing InSe flakes of various thicknesses ---in between 6 and 8 layers--- on planar (\emph{i.e.}, strain-free) MoS$_2$ (\textbf{a}-\textbf{b}) or WS$_2$ (\textbf{c}-\textbf{d}) MLs. A decrease of the InSe PL signal in the HS region (IN) with respect to the bare InSe flake (OUT) is observed in all cases, contrary to the PL enhancement observed in the strained case.
        %


	\newpage
	\section*{
        Supporting Note 5. Photoluminescence excitation studies of an unstrained hBN/InSe/MoS$_2$ heterostructure
        }
	\addcontentsline{toc}{section}{
        Supporting Note 5. Photoluminescence excitation studies of an unstrained hBN/InSe/MoS$_2$ heterostructure
        }
	\setcounter{section}{5}
        \sectionmark{}
	\setcounter{figure}{0}
	\setcounter{equation}{0}
        \setcounter{table}{0}

\begin{figure*}[htbp]
        \begin{center} \includegraphics[width=0.4\textwidth]{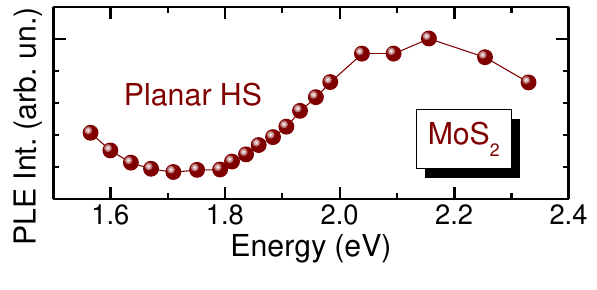}
        \caption{
        \textbf{Photoluminescence excitation measurement of a planar heterostructure}. PLE spectrum of the hBN/6L-InSe/MoS$_2$ planar HS of Fig.\ \ref{fig:check_planar}. 
        }
        \label{fig:PLE_planar}
        \end{center}
        \end{figure*}

        Fig.\ \ref{fig:PLE_planar} shows the PLE spectrum of the hBN/6L-InSe/MoS$_2$ planar HS whose PL spectrum is shown in Fig.\ \ref{fig:check_planar}. The PLE spectrum was acquired by monitoring the PL signal of InSe while the excitation laser energy was changed. Indeed, at variance with the PLE spectrum of the strained HS bubble (shown in Fig.\ 4 of the main text), in this case no clear resonance is observed, suggesting that no charge transfer from the MoS$_2$ ML to the InSe flake occurs. This result is consistent with the absence of PL enhancement observed in our planar HSs (see \textcolor{purple}{Supporting Note 4}) and in Ref.\ \cite{funneling_MoS2InSe}.

        %


	\newpage
	\section*{
        Supporting Note 6. DFT simulations of explicit InSe/MoS$_2$ interfaces.
        }
	\addcontentsline{toc}{section}{
        Supporting Note 6. DFT simulations of explicit InSe/MoS$_2$ interfaces.
        }
	\setcounter{section}{6}
        \sectionmark{}
	\setcounter{figure}{0}
	\setcounter{equation}{0}
        \setcounter{table}{0}

\begin{figure}[h]
\centering
\includegraphics[width=.7\textwidth]{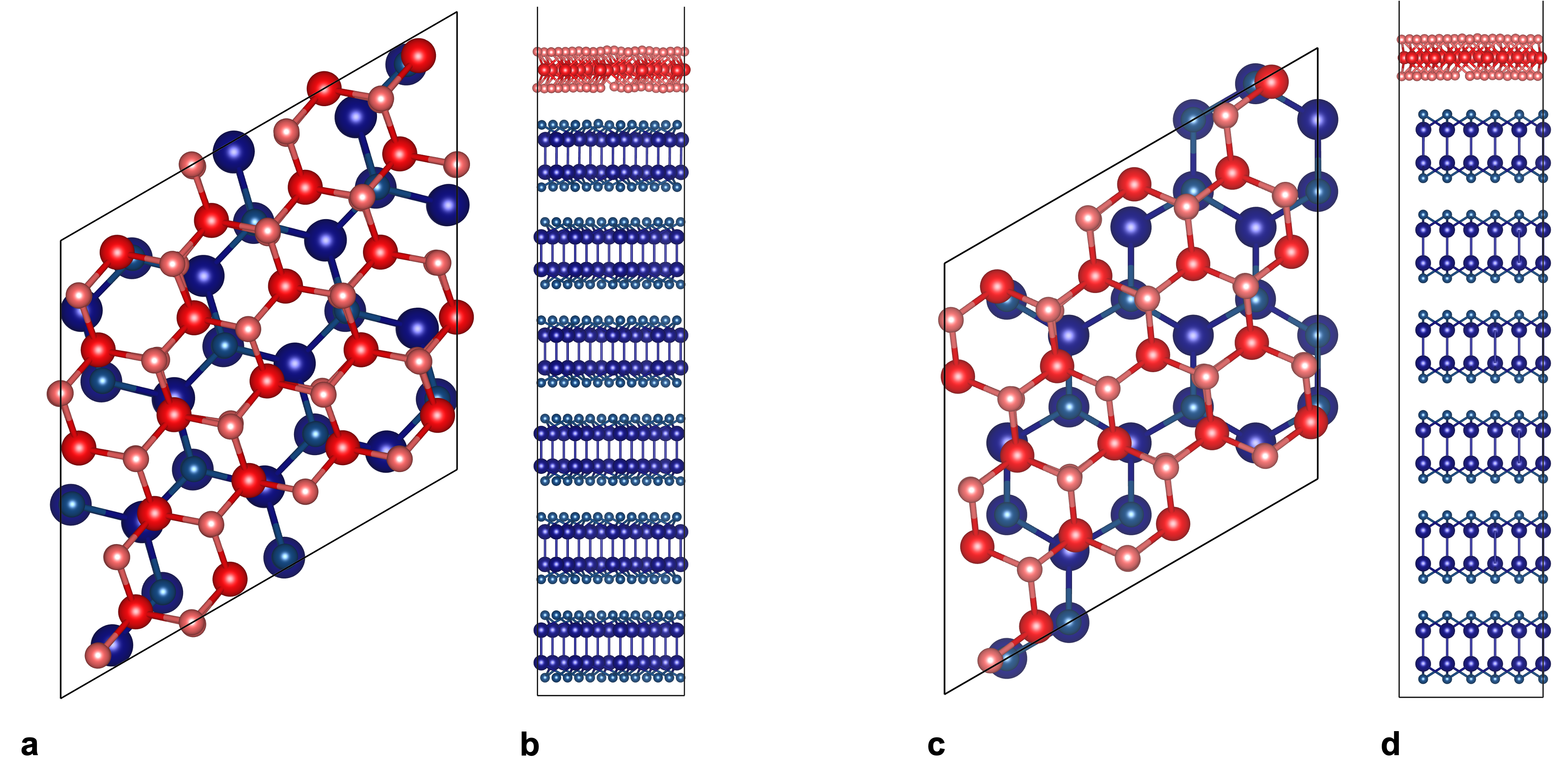}
\caption{Optimised geometries of explicit HSs made of 6L-InSe and MoS$_2$ ML with sulphur vacancy. \textbf{a}-\textbf{b}: top and side view of the HS with the unstrained MoS$_2$ ML, respectively. \textbf{c}-\textbf{d}: top and side view of the HS with MoS$_2$ ML at 2\% biaxial tensile strain, respectively. Large dark red and smaller light red spheres mark Mo and S atoms, respectively. Large dark blue and smaller light blue spheres mark In and Se atoms, respectively.}
\label{fig:geometries_HS}
\end{figure}

We built explicit HSs between 6L-InSe and ML-MoS$_2$ with a sulphur vacancy. While 6L-InSe was unstrained, the TMD ML was studied both at 0\% and at 2\% biaxial tensile strain. Side and top view of both HSs are shown in Figure~\ref{fig:geometries_HS}. 

The unstrained and strained HSs count 3504 and 3224 electrons, respectively. All calculations, from geometry optimisation to band structure and projected density of states (pDOS) evaluation were performed with the GGA-based rev-vdW-DF2 functional \cite{vdW-DF2} without spin-orbit coupling. Band structures and $k$-resolved pDOS ($k$-pDOS) are shown for $k$-points in the Brillouin zone of the relative HS.

\begin{figure}[h]
\centering
\includegraphics[width=.8\textwidth]{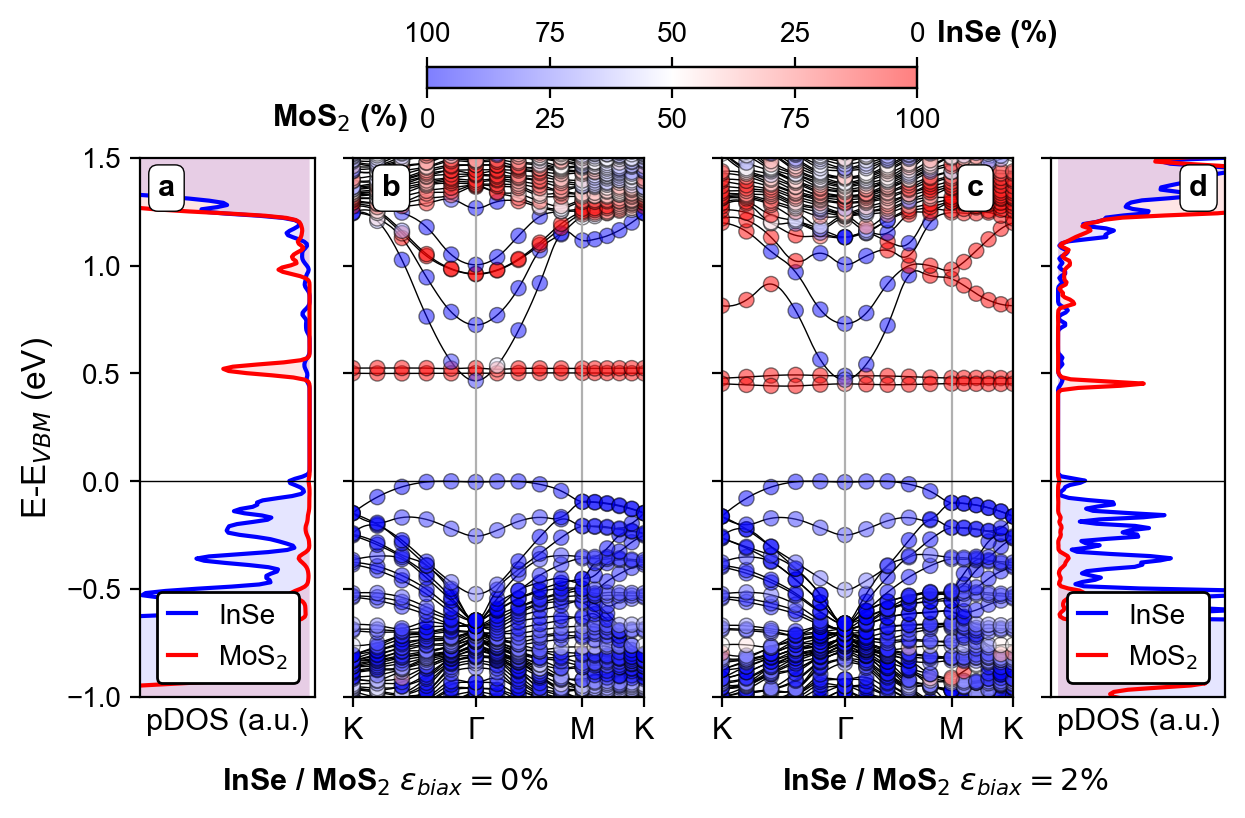}
\caption{Band structures, $k$-pDOS and pDOS of the HSs made of 6L-InSe and MoS$_2$ ML with a sulphur vacancy. \textbf{a}-\textbf{b} Unstrained MoS$_2$ ML case. \textbf{c}-\textbf{d} 2\% strained MoS$_2$ ML case.}
\label{fig:explicit_HS}
\end{figure}

In Figure~\ref{fig:explicit_HS} we report the band structures with superimposed the $k$-resolved density of states projected on the atomic obitals of 6L-InSe and ML-MoS$_2$. Red and blue dots corresponds to electronic states composed of MoS$_2$ or InSe atomic orbitals, respectively. Intermediate shades mark hybridization. Notice that in Figure~\ref{fig:explicit_HS}\textbf{b}, the K/K' points of MoS$_2$ are folded to the $\Gamma$ point of the HS Brillouin zone. 

\begin{figure}[h]
\centering
\includegraphics[width=.8\textwidth]{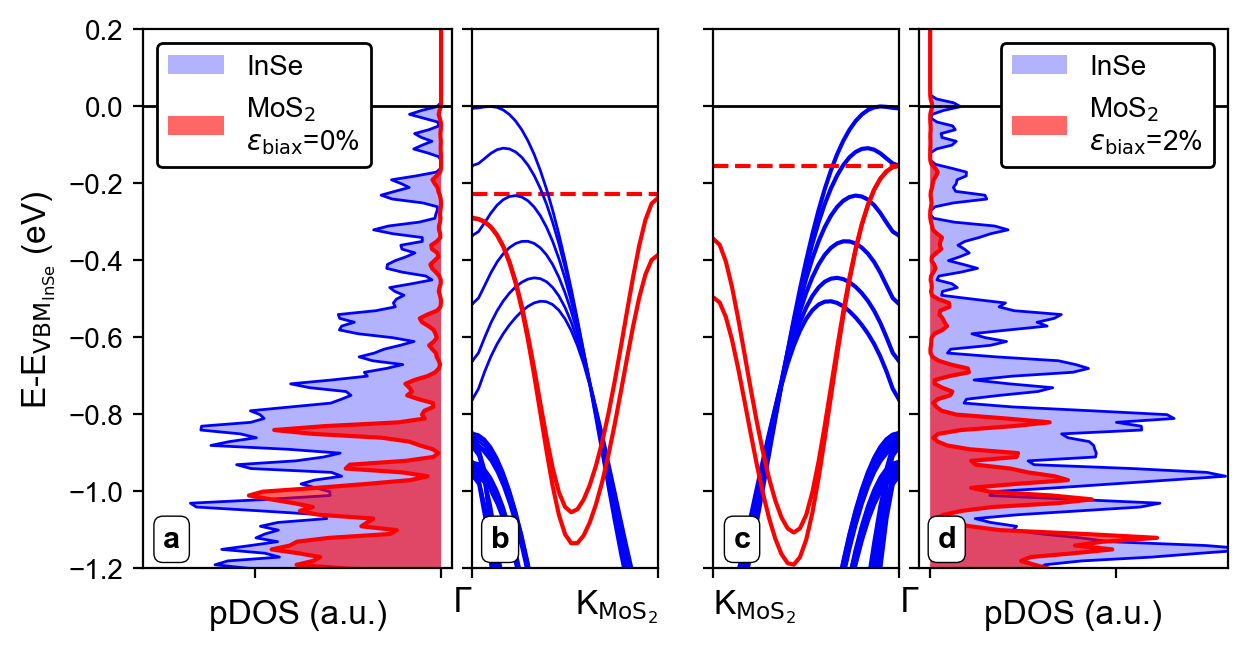}
\caption{pDOS of the explicit HSs with unstrained, (panel \textbf{a}), and strained (panel \textbf{d}) ML-MoS$_2$. In panels \textbf{b} and \textbf{c} we report the relative valence band alignment as predicted from GGA calculations via the method in Ref.~\cite{Kresse}.}
\label{fig:pdos_HS}
\end{figure}

In Figure~\ref{fig:pdos_HS} we report a detail of the pDOS of 6L-InSe and ML-MoS$_2$ in the explicit HSs for $\varepsilon_\textup{biax}=0\%$ of MoS$_2$, in panel \textbf{a} and $\varepsilon_\textup{biax}=2\%$ in \textbf{d}. We compare the pDOS obtained from the full simulations of the HSs to the aligned band structures of both materials (panels \textbf{b} and \textbf{c}), in their unit cells. The band alignment in panels \textbf{b} and \textbf{c} is obtained from PBE calculations, independently of the explicit HSs simulations, and following the method in Ref.~\cite{Kresse}, used for the aligment of the HSE band structures in the main text.

The comparison between the reconstructed alignments in panels \textbf{b} and \textbf{c} and the relative pDOSs in panels \textbf{a} and \textbf{b}, respectively, further confirms that the top valence band of MoS$_2$ at the $\Gamma$ point aligns with the topmost valence bands of 6L-InSe. In particular, the VBM of the strained MoS$_2$ monolayer, at $\Gamma$, shifts closer to the VBM of InSe.


 \clearpage

\newpage

\renewcommand{\bibname}{References}
\addcontentsline{toc}{section}{References}

\bibliographystyle{thesisamj}
\bibliography{jabbr,mybib}

\end{document}